\documentclass[a4paper,11pt]{article}
\usepackage{jheppub} 
\usepackage{lineno}

\usepackage{graphicx} 
\usepackage[T1]{fontenc}
\usepackage{amsmath}
\usepackage{graphicx}
\usepackage{latexsym,amssymb,lmodern}
\usepackage{float}
\usepackage{cleveref}
\usepackage[colorinlistoftodos]{todonotes}
\usepackage{hyperref}
\usepackage{xcolor}
\usepackage{soul}
\usepackage{cancel}
\usepackage{appendix}

\usepackage{wrapfig}
\usepackage{braket}
\usepackage{physics}
\usepackage{ulem}
\usepackage{cancel}
\usepackage{comment}
\usepackage{enumitem}
\usepackage{slashed}
\usepackage{mathscinet}
\usepackage{dsfont} 
\usepackage{simpler-wick}
\usepackage{fontawesome5}
\usepackage{xspace}
\usepackage{mathtools}
\usepackage{tikz}

\usetikzlibrary{decorations.pathmorphing}
\usetikzlibrary{decorations.pathreplacing}
\usetikzlibrary{arrows.meta,decorations.markings}
\usetikzlibrary{patterns}
\usetikzlibrary{intersections}
\usetikzlibrary{calc,shapes.misc,bending}

\tikzset{
  loopprop/.style={
    black, very thick, line cap=round, line join=round
  },
  axionline/.style={
    black, very thick, dash pattern=on 6pt off 5pt, line cap=round
  },
  gravitonwave/.style={
    black, very thick,
    decorate,
    decoration={snake, amplitude=1.8pt, segment length=7pt}
  }
}

\newcommand{\LoopArrowLine}[3][]{%
  \draw[
    loopprop,#1,
    postaction={decorate},
    decoration={markings,
      mark=at position 0.58 with {\arrow{Latex[length=3mm,width=2mm]}}
    }
  ] (#2) -- (#3);
}

\newcommand{\GravitonLine}[2]{%
  \draw[gravitonwave]
    ($(#1)!1.2pt! 90:(#2)$) -- ($(#2)!1.2pt!-90:(#1)$);
  \draw[gravitonwave]
    ($(#1)!1.2pt!-90:(#2)$) -- ($(#2)!1.2pt! 90:(#1)$);
}

\newcommand{\CircleAHHLoop}[1][1]{%
\begin{tikzpicture}[scale=#1, line cap=round, line join=round, >=Latex]
  \coordinate (C) at (0,0);
  \def\r{1.00}

  \coordinate (L) at (-3.0,0);
  \coordinate (Lin) at (-\r,0);  
  \draw[axionline] (L) -- (Lin);

\draw[
    loopprop,
    postaction={decorate},
    decoration={markings,
      mark=at position 0.3 with {
        \pgftransformrotate{12}
        \arrow{Latex[length=2.3mm,width=2.3mm]}
      },
      mark=at position 0.8 with {
        \pgftransformrotate{12}
        \arrow{Latex[length=2.3mm,width=2.3mm]}
      }
    }
  ] ($(C)+(\r,0)$) arc[start angle=0,end angle=-360,radius=\r];

  \coordinate (Vt) at ($(C)+(4:\r)$);
  \coordinate (Vb) at ($(C)+(-4:\r)$);

  \coordinate (Gt) at (3.2,1.7);
  \coordinate (Gb) at (3.2,-1.7);
  \GravitonLine{Vt}{Gt}
  \GravitonLine{Vb}{Gb}

  \node[font=\Large, anchor=east] at ($(L)+(-0.10,0.20)$) {$\phi$};
  \node[font=\Large, anchor=west] at ($(Gt)+(0.15,0.15)$) {$h_{ab}$};
  \node[font=\Large, anchor=west] at ($(Gb)+(0.15,-0.05)$) {$h_{cd}$};
\end{tikzpicture}%
}

\newcommand{\TriangleAHHLoop}[1][1]{%
\begin{tikzpicture}[scale=#1, line cap=round, line join=round, >=Latex]
  \coordinate (L) at (-2.4,0);
  \coordinate (T) at ( 0.8,1.6);
  \coordinate (B) at ( 0.8,-1.6);

  \coordinate (Lin) at (-4.1,0);
  \draw[axionline] (Lin) -- (L);

  \LoopArrowLine{T}{L}
  \LoopArrowLine{L}{B}
  \LoopArrowLine{B}{T}

  \coordinate (Gt) at (3.2,1.6);
  \coordinate (Gb) at (3.2,-1.6);
  \GravitonLine{T}{Gt}
  \GravitonLine{B}{Gb}

  \node[font=\Large, anchor=east] at ($(Lin)+(-0.10,0.20)$) {$\phi$};
  \node[font=\Large, anchor=west] at ($(Gt)+(0.15,0.15)$) {$h_{ab}$};
  \node[font=\Large, anchor=west] at ($(Gb)+(0.15,-0.05)$) {$h_{cd}$};
\end{tikzpicture}%
}

\newcommand{\BoxAAtoHHLoop}[1][1]{%
\begin{tikzpicture}[scale=#1, line cap=round, line join=round, >=Latex]
  \coordinate (TL) at (-1.2, 1.4);
  \coordinate (TR) at ( 1.2, 1.4);
  \coordinate (BR) at ( 1.2,-1.4);
  \coordinate (BL) at (-1.2,-1.4);

  \coordinate (AL1) at (-4.1, 2.0);
  \coordinate (AL2) at (-4.1,-2.0);
  \draw[axionline] (AL1) -- (TL);
  \draw[axionline] (AL2) -- (BL);

  \LoopArrowLine{TR}{TL}
  \LoopArrowLine{TL}{BL}
  \LoopArrowLine{BL}{BR}
  \LoopArrowLine{BR}{TR}

  \coordinate (Gt) at (4.1, 1.4);
  \coordinate (Gb) at (4.1,-1.4);
  \GravitonLine{TR}{Gt}
  \GravitonLine{BR}{Gb}

  \node[font=\Large, anchor=east] at ($(AL1)+(-0.10,0.20)$) {$\phi$};
  \node[font=\Large, anchor=east] at ($(AL2)+(-0.10,-0.20)$) {$\phi$};
  \node[font=\Large, anchor=west] at ($(Gt)+(0.15,0.15)$) {$h_{ab}$};
  \node[font=\Large, anchor=west] at ($(Gb)+(0.15,-0.05)$) {$h_{cd}$};
\end{tikzpicture}%
}

\newcommand{\SChannelAxion}[1][1]{%
\begin{tikzpicture}[baseline={([yshift=-0.5ex]0,0)}, scale=#1, line cap=round, line join=round, >=Latex]
  \coordinate (VL) at (-1.5, 0);
  \coordinate (VR) at ( 1.5, 0);

  \coordinate (In1) at (-3.0,  1.5);
  \coordinate (In2) at (-3.0, -1.5);
  \coordinate (Out1) at (3.0,  1.5);
  \coordinate (Out2) at (3.0, -1.5);

  \LoopArrowLine{In1}{VL}
  \LoopArrowLine{In2}{VL}
  \LoopArrowLine{VR}{Out1}
  \LoopArrowLine{VR}{Out2}

  \draw[axionline] (VL) -- (VR);

  \node[font=\normalsize, anchor=east] at ($(In1)+(-0.15, 0.15)$) {$\psi$};
  \node[font=\normalsize, anchor=east] at ($(In2)+(-0.15,-0.15)$) {$\psi$};
  \node[font=\normalsize, anchor=west] at ($(Out1)+(0.15, 0.15)$) {$\psi$};
  \node[font=\normalsize, anchor=west] at ($(Out2)+(0.15,-0.15)$) {$\psi$};
\end{tikzpicture}%
}

\newcommand{\SChannelAxionLoop}[1][1]{%
\begin{tikzpicture}[baseline={([yshift=-0.5ex]0,0)}, scale=#1, line cap=round, line join=round, >=Latex]
  \coordinate (VL) at (-2.3, 0);
  \coordinate (VR) at ( 2.3, 0);

  \coordinate (C) at (0,0);
  \def\r{0.8}
  \coordinate (BL) at (-\r, 0);
  \coordinate (BR) at ( \r, 0);

  \coordinate (In1) at (-3.8,  1.5);
  \coordinate (In2) at (-3.8, -1.5);
  \coordinate (Out1) at (3.8,  1.5);
  \coordinate (Out2) at (3.8, -1.5);

  \LoopArrowLine{In1}{VL}
  \LoopArrowLine{In2}{VL}
  \LoopArrowLine{VR}{Out1}
  \LoopArrowLine{VR}{Out2}

  \draw[axionline] (VL) -- (BL);
  \draw[axionline] (BR) -- (VR);

  \draw[
    loopprop,
    postaction={decorate},
    decoration={markings,
      mark=at position 0.25 with {
        \pgftransformrotate{12}
        \arrow{Latex[length=2.5mm,width=2.5mm]}
      },
      mark=at position 0.75 with {
        \pgftransformrotate{12}
        \arrow{Latex[length=2.5mm,width=2.5mm]}
      }
    }
  ] ($(C)+(\r,0)$) arc[start angle=0,end angle=-360,radius=\r];

  \node[font=\normalsize, anchor=east] at ($(In1)+(-0.15, 0.15)$) {$\psi$};
  \node[font=\normalsize, anchor=east] at ($(In2)+(-0.15,-0.15)$) {$\psi$};
  \node[font=\normalsize, anchor=west] at ($(Out1)+(0.15, 0.15)$) {$\psi$};
  \node[font=\normalsize, anchor=west] at ($(Out2)+(0.15,-0.15)$) {$\psi$};
\end{tikzpicture}%
}

\newcommand{\SChannelGraviton}[1][1]{%
\begin{tikzpicture}[baseline={([yshift=-0.5ex]0,0)}, scale=#1, line cap=round, line join=round, >=Latex]
  \coordinate (VL) at (-1.5, 0);
  \coordinate (VR) at ( 1.5, 0);

  \coordinate (In1) at (-3.0,  1.5);
  \coordinate (In2) at (-3.0, -1.5);
  \coordinate (Out1) at (3.0,  1.5);
  \coordinate (Out2) at (3.0, -1.5);

  \LoopArrowLine{In1}{VL}
  \LoopArrowLine{In2}{VL}
  \LoopArrowLine{VR}{Out1}
  \LoopArrowLine{VR}{Out2}

  \GravitonLine{VL}{VR}

  \node[font=\normalsize, anchor=east] at ($(In1)+(-0.15, 0.15)$) {SM};
  \node[font=\normalsize, anchor=east] at ($(In2)+(-0.15,-0.15)$) {SM};
  \node[font=\normalsize, anchor=west] at ($(Out1)+(0.15, 0.15)$) {SM};
  \node[font=\normalsize, anchor=west] at ($(Out2)+(0.15,-0.15)$) {SM};
\end{tikzpicture}%
}

\newcommand{\SChannelGravitonLoop}[1][1]{%
\begin{tikzpicture}[baseline={([yshift=-0.5ex]0,0)}, scale=#1, line cap=round, line join=round, >=Latex]
  \coordinate (VL) at (-2.4, 0);
  \coordinate (VR) at ( 2.4, 0);

  \coordinate (C) at (0,0);
  \def\r{0.8}
  \coordinate (BL) at (-\r, 0);
  \coordinate (BR) at ( \r, 0);

  \coordinate (In1) at (-3.9,  1.5);
  \coordinate (In2) at (-3.9, -1.5);
  \coordinate (Out1) at (3.9,  1.5);
  \coordinate (Out2) at (3.9, -1.5);

  \LoopArrowLine{In1}{VL}
  \LoopArrowLine{In2}{VL}
  \LoopArrowLine{VR}{Out1}
  \LoopArrowLine{VR}{Out2}

  \GravitonLine{VL}{BL}
  \GravitonLine{BR}{VR}

  \draw[
    loopprop,
    postaction={decorate},
    decoration={markings,
      mark=at position 0.25 with {
        \pgftransformrotate{12}
        \arrow{Latex[length=2.5mm,width=2.5mm]}
      },
      mark=at position 0.75 with {
        \pgftransformrotate{12}
        \arrow{Latex[length=2.5mm,width=2.5mm]}
      }
    }
  ] ($(C)+(\r,0)$) arc[start angle=0,end angle=-360,radius=\r];

  \node[font=\normalsize, anchor=east] at ($(In1)+(-0.15, 0.15)$) {SM};
  \node[font=\normalsize, anchor=east] at ($(In2)+(-0.15,-0.15)$) {SM};
  \node[font=\normalsize, anchor=west] at ($(Out1)+(0.15, 0.15)$) {SM};
  \node[font=\normalsize, anchor=west] at ($(Out2)+(0.15,-0.15)$) {SM};
\end{tikzpicture}%
}

\makeatletter
\newcommand{\github}[1]{%
   \href{#1}{\faGithubSquare}%
}
\makeatother

\definecolor{blue3}{RGB}{31,119,180}
\definecolor{red3}{RGB}{214,39,40}
\definecolor{orange3}{RGB}{255,127,14}
\definecolor{green3}{RGB}{44,160,44}

\newcommand{\ac}[1]{\textcolor{green3}{[\textbf{AC:}] #1}}

\newcommand{\com}[1]{\textcolor{red}{[\textbf{Comment:}] #1}}

\newcommand{\mpl}{M_{\rm Pl}}
\newcommand{\h}{\overline{h}}
\newcommand{\g}{\overline{g}}
\newcommand{\szeta}{\alpha}
\newcommand{\HH}{\hat{H}}
\newcommand{\factor}{f}
\newcommand{\Lamc}{\Lambda_\text{causality}}
\newcommand{\Lams}{\Lambda_\text{species}}

\numberwithin{equation}{section}

\newmuskip\pFqskip
\mathchardef\pFcomma=\mathcode`, 

\title{Theoretical and Observational Bounds on Dynamical Chern-Simons Gravity as an Effective Field Theory}

\author{Alexander Cassem and Mark P.~Hertzberg}
\affiliation{Institute of Cosmology, Department of Physics and Astronomy, Tufts University, Medford, MA 02155, USA}

\emailAdd{alexander.cassem@tufts.edu}
\emailAdd{mark.hertzberg@tufts.edu}

\abstract{
Gravitational effective theories are essential for characterizing the space of  deviations from General Relativity (GR). 
Testing these theories against fundamental principles, such as causality and unitarity, can yield constraints on the corresponding parameters. In this paper, we perform such an analysis on the very interesting dynamical Chern-Simons (dCS) gravity. This is a parity violating correction to GR wherein a new scalar field couples to the Pontryagin density $^*R\,R$. It has generated significant interest, including possible new gravitational wave shapes for LIGO/Virgo and new phenomena from cosmic inflation.
In this work, we begin by deriving the dispersion relation and wave packet speed on top of a gravitational wave background in dCS gravity. This alters the corresponding Shapiro time delay (which we compute to second order), potentially giving superluminality. Causality then demands a bound on the dCS coupling constant, which we find to be moderately sharper than, but compatible with, standard estimates. 
We then examine a UV completion in the form of a set of $N$ fermions with a (pseudo) Yukawa coupling. 
By imposing perturbativity and a gravitational species bound, we find that the dCS coupling constant is constrained significantly more, depending on the choice of scale of the species bound.
We also identify higher order operators generated from the UV completion.
Overall, we find that any dCS corrections to gravitational dynamics should likely be very small on macroscopic systems of observational interest, such as in late-time merging black holes.
}

\begin{document}
\maketitle
\flushbottom

\newpage
\section{Introduction}\label{sec: introduction}

Within the Standard Model of particle physics, the electroweak sector maximally violates parity (originally seen in experiments such as in cobalt-60 \cite{LeeYang1956,Wu:1957my}). While the strong sector has yet to show parity violation, despite the theoretical possibility of the $\theta\,G\,\widetilde{G}$ (with tight constraints on the electric dipole moment of the neutron \cite{Baker:2006ts,Abel:2020gbr}). 
It is important to ask what other sectors of physics beyond the Standard Model may also give rise to parity violation. 

If we turn from particle physics  towards cosmology, we can ask if there are signals of parity violation in that context. 
Intriguingly, for some time, there were suggestions that there is observed parity violation in the large scale structure \cite{Coulton:2023oug,Diego-Palazuelos:2022cnh,Hou:2022wfj,Minami:2020odp,Philcox:2022hkh}.
This came from detailed analyses of the four-point correlation function (trispectrum) of the Large Scale Structure (LSS). Models of this primordial parity violation from inflation were put forward \cite{Creque-Sarbinowski:2023wmb,Fujita:2023AxionTrispectrum,Stefanyszyn:2025yhq}.
However, recently this evidence has reduced \cite{Krolewski:2024paz}.  

On the theoretical side of cosmological parity violation, there is a proposed ``no-go theorem'' for parity violation during inflation \cite{Liu:2019fag,Cabass:2022rhr,Thavanesan:2025kyc}. Crucially this focuses on standard General Relativity (GR), rather than incorporating large corrections. In Ref.~\cite{Creque-Sarbinowski:2023wmb} a very interesting proposal for obtaining significant parity violation in the inflationary trispectrum was put forward by \textit{directly} modifying gravity. 

This made use of  \textit{dynamical Chern-Simons} (dCS) gravity (see \cite{Jackiw:2003pm,Alexander:2009tp} for details on formulation and consequences of dCS gravity). This is a theory in which a new light scalar $\phi$ is assumed to couple to the Pontryagin index as $\phi\,^*R\,R$. It classically carries a shift symmetry and therefore the lightness of $\phi$ is justifiable. As an effective field theory (EFT), it is an interesting correction to GR as it only involves 4 derivatives. This should be compared to pure gravity corrections of the form $R^3$, $R^4$ (where this is shorthand for contractions of the Riemann tensor), which are 6 and 8 derivatives, respectively. (Note that any $R^2$ term is a total derivative in 4-dimensions, Gauss-Bonnet or $^*R\,R$, so it does not contribute at the classical level). So from a power counting point of view, dCS is a kind of leading order correction to GR.

Within the past few years, a plethora of theoretical work on constraining the Wilson coefficients for the EFT of GR has been performed.
In particular, there are established restrictions on the Wilson coefficients of infrared (IR) operators in order for the ultraviolet (UV) theory to have analyticity. This is closely related to causality \cite{Adams:2006sv}. This has led to an abundance of constraints on the Wilson coefficients of higher order curvature operators, see for example \cite{Herrero-Valea:2022lfd,Platania:2022gtt,deRham:2022gfe,Hong:2023zgm,Bellazzini:2015cra,Hamada:2023cyt}. 
This includes bounds on the $R^3$ and $R^4$ coefficients \cite{Gruzinov:2006ie,Camanho:2014apa,Caron-Huot:2022ugt}.  
These inequalities on the Wilson coefficients can be used in combination with experimental constraints to guide searches of extensions of GR \cite{Horowitz:2024dch,Cassem:2024djm,Hertzberg:2025heq}. 
This begs the question: are there constraints from fundamental principles such as causality, locality, and unitarity, that already put significant constraints on  dynamical Chern-Simons?

In this work we will mainly be interested in the possible consequences of the dCS operator on macroscopic systems in the late universe, especially black holes or neutron stars, of the sort LIGO/Virgo is sensitive to \cite{Abbott:2016blz,Abbott:2016nmj,Abbott:2017oio}. Interesting work on observational 
bounds on dCS is in Refs.~\cite{Yunes:2016jcc,Okounkova:2017zgw,Perkins:2021mhb,Chung:2025dcs,Yunes:2013dva,Srivastava:2021imr,Falkowski:2024bgb,Falkowski:2024yuy,Silva:2020acr,Silva:2022srr}. 
We will primarily use theoretical arguments to determine  the precision with which dCS could potentially alter the gravitational dynamics and gravitational wave signal in this context. We will leave possible consequences for inflation and the very early universe for future analyses.

Interestingly, there has also been work on deriving the Shapiro time delay and deflection angle from the eikonal approximation of the S-matrix, see for example \cite{Kabat:1992tb,Amati:1992zb,AccettulliHuber:2020oou}. In particular, in \cite{Serra:2022pzl}, they looked at the time delay from the dCS term due to a static point source, and used this to place a bound on the corresponding Wilson coefficient. 
Other work using amplitudes appears in Refs.~\cite{Xu:2024pvb,Dong:2025dpy}.
On the other hand, the use of causality and weak gravity arguments have been suggested in Ref.~\cite{Alexander:2025gdn} to be rather unhelpful in constraining modifications of gravity. 

In this work, we compute the time delay of a wave packet on a novel background: a gravitational wave. We find that we need to work to second order in the GR contribution, while we can work to first order from the dCS contribution.
(An example can be a binary black hole system that generates the background gravitational wave.)
We find a bound on the dCS coupling depending on how far into the UV one can push the frequency of the wave packet that rides on top of the background. If we do not push this frequency too much higher than the causality cutoff itself, then the causality bound is parametrically of the same order as the (inverse) length scale set by the dCS Wilson coefficient.

Furthermore, we carefully investigate a UV completion of dCS from the scalar with a pseudo-Yukawa coupling to a set of massive fermions. We allow for $N$ fermions for maximal freedom in the completion. We show that the requirements of perturbativity and a gravitational species bound provides a further constraint on the dCS coefficient. In fact, if we demand the species bound cutoff is above already probed scales, then the dCS operator is constrained to be extremely small on reasonable macroscopic scales. 

The rest of the paper is organized as follows: 
In Section \ref{sec: Dyanmical Chern-Simons Gravity} we derive the equations of motion from the EFT. We then perturb around some background spacetime. We then derive the dispersion relation for the perturbation, and confirm it matches with that found in \cite{Garfinkle:2010zx,Ayzenberg:2013wua}. 
In Section \ref{sec: Shapiro Time Delay on a Gravitational Wave Background}, we specify the background to be that of a gravitational wave. Working to second order, we derive the Shapiro time delay and the dCS advance for a special eigen-mode. 
This leads to a constraint on the coupling.
In Section \ref{sec: Constraints from Unitarity, and UV Completions}, we study the UV completion from fermions and derive an improved bound.
We discuss the implications of our results for observations. 
Finally, we discuss our results in Section \ref{sec: conclusions and discussion}. 


We include four appendices. In Appendix \ref{sec: appendix for perturbed eom and consistency}, we explicitly derive the dispersion relation used in this paper, as well as show consistency in the equations of motion based on the helicity of the spin-2 graviton similar to that found in the geometric optics literature \cite{Shore:2002gn,Shore:2007um}, which was missing in the literature. In Appendix \ref{app: shapiro expressions appendix}, we state explicitly the Shapiro time delay at second order. In Appendix \ref{app: shapiro time delay near a black hole appendix}, we investigate what happens when the setup for the time delay is close to a black hole, and picks up a redshift factor.  Finally, in Appendix \ref{app: UV Completion of dCS and EFT via Heat Kernel}, we give an overview of how to generate the dCS term from a chiral-rotation, as well as a discussion on higher dimension operators that originate from integrating out  massive chiral fermions.

\textbf{Notations and computer use:} Throughout this paper we work in the mostly plus metric, $(-,+,+,+)$ sign convention. We set $c = \hbar  = 1$.  Any quantity with a bold face such as $\mathbf{k}$ denotes the spatial components of the given quantity. Our metric perturbation and scalar perturbation are written as
\begin{align}
    g_{ab}& = \overline{g}_{ab} + h_{ab}\label{eq: metric perturbation def}\\
    \phi & = \overline{\phi} + \delta\phi\label{eq: phi perturbation def}
\end{align}
where any quantity with a ``bar'' such as $\overline{R}_{ab}$ denotes the quantity is constructed with respect to the background metric $\overline{g}_{ab}$, while $h_{ab}$ and $\delta\phi$ represent perturbations. 

The majority of the tensor calculations and manipulations calculated here have been aided by the usage of the \textit{Mathematica} package \textit{xAct} (see this webpage for more information: \href{https://www.xact.es/}{xAct}) as well as multiple sub-packages \cite{Martin-Garcia:2007bqa,Martin-Garcia:2008yei,Martin-Garcia:2008ysv,Brizuela:2008ra,Nutma:2013zea}. A \textit{Mathematica} notebook with the majority of  derivations performed in this paper is provided, and can be found at the following link \href{https://github.com/acassem/Dynamical-Chern-Simons-Time-Delay}{\textcolor{violet}{\github{}}}.

\section{Dynamical Chern-Simons Gravity}\label{sec: Dyanmical Chern-Simons Gravity}

In this section, we briefly review the theory of dCS gravity while deriving the equations of motion, and then derive the cutoff of dCS from unitarity. We then perturb the equations of motion on a general background, and then specialize to the case of when $\overline{R}_{ab} = 0$ which simplifies them allowing us to derive the corresponding dispersion relation.

\subsection{Action and Equations of Motion}

The full action we consider is of the form
\begin{equation}
    S = S_{\text{EH}} + S_\text{dCS} + S_\phi + S_\text{matter}\label{eq: total action}
\end{equation}
Here the first term is the Einstein-Hilbert action of GR
\begin{equation}
    S_\text{EH} = \int d^4x\sqrt{-g}\left[\mpl^2\,R\right]\label{eq: einstein hilbert action}
\end{equation}
where $R$ is the Ricci scalar and $\mpl = 1/\sqrt{16\pi G}$, with $G$ Newton's gravitational constant and $\mpl$ is a (reduced) Planck mass. It is the unique term (modulo a cosmological constant) governing the dynamics of a massless spin 2 particle at large distances.

The gravitational dynamics can only be corrected by adding higher dimension operators and/or adding new degrees of freedom. Here we include a scalar $\phi$ with action 
\begin{equation}
    S_\phi = -\int d^4x\sqrt{-g}\left[{1\over2}g^{ab}\nabla_a\phi\nabla_b\phi + V(\phi)\right]\label{eq: massless scalar action}.
\end{equation}
A generic scalar can be expected to be massive, and then it will not have long ranged consequences. For a massless scalar, we should endow its action with a shift symmetry (or approximate shift symmetry). So we shall soon set $V=0$.

We introduce new gravitational dynamics by way of the 
gravitational Chern-Simons term
\begin{equation}
    S_\text{dCS} = \int d^4x\sqrt{-g}\left[{\alpha\,\mpl\over4}\,\phi\, ^*R\,R\right]
    \label{eq: gravitational chern simons action}
\end{equation}
where $^*R\,R$ is the Pontryagin density with the dual of the Riemann tensor defined as\footnote{The dual of the Riemann tensor is taken on the last 2 indices of the Riemann tensor.}
\begin{equation}
    {}^*R^a{}_b{}^{cd}\equiv \frac{1}{2}\epsilon^{cdef}R^a{}_{bef}\label{eq: dual riemann tensor}.
\end{equation}
Here, $\alpha$ is the dCS coupling constant with units 
\begin{equation}
    [\alpha]=(\mbox{length})^2.
    \end{equation}
    It sets the characteristic length scale over which this dCS operator has an impact in the strong gravity regime.
Since $\sqrt{-g}\,^*R\,R$ can be shown to be a total derivative, then this dCS action carries the shift symmetry
\begin{equation}
    \phi\to\phi+\mbox{const}
\end{equation}
at the classical level. At the quantum level, it is shift symmetric, up to a topological term. In either case, this provides a reasonable justification for the masslessness of $\phi$, as we shall assume.

Finally, the fourth term in \eqref{eq: total action} is a generic matter contribution we can write as
\begin{equation}
    S_\text{matter} = \int d^4x\sqrt{-g}\,\mathcal{L}_\text{matter}\label{eq: matter action}
\end{equation}
where $\mathcal{L}_\text{matter}$ is assumed to contain no dependence on $\phi$. So $\phi$ is decoupled from the matter (Standard Model) sector.

By varying the action of \eqref{eq: total action} with respect to the metric $g_{ab}$ and the scalar field $\phi$, we have the following pair of equations of motion
\begin{align}
    G^{ab}  + {\alpha\over\mpl}\, C^{ab} & = {1\over2\mpl^2}T^{ab}\label{eq: metric eom}\\
    \nabla_a\nabla^a\phi - V'(\phi) & = - \frac{\alpha\,\mpl}{4}\,{}^*R_a{}^{bcd}R^a{}_{bcd} \label{eq: scalar eom}
\end{align}
where $G^{ab}$ is the Einstein tensor, $T^{ab}$ is the scalar energy-momentum tensor
\begin{equation}
    T^{ab} = 
    \nabla^a\phi\nabla^b\phi - \frac{1}{2}g^{ab}\nabla_c\phi\nabla^c\phi 
     - g^{ab}V(\phi)\label{eq: scalar energy momentum tensor},
\end{equation}
and $C^{ab}$ is the so-called \textit{C-tensor} which is defined as\footnote{The symmetrization operation is defined as $A^{(ab)}\equiv \frac{1}{2}(A^{ab} + A^{ba})$.}
\begin{equation}
    C^{ab} = {}^*R^{d(ab)c}\nabla_e\nabla_d\phi + \left(\nabla_c\phi\right)\epsilon^{cde(b}\nabla_eR^{a)}{}_d\label{eq: c-tensor}.
\end{equation}

\subsection{Unitarity}\label{subsec: unitarity sec 2}

The dynamical Chern-Simons term is a higher dimension operator which leads to a low cutoff on the theory.

To establish this cutoff, let us consider 2-to-2 graviton scattering via scalar exchange. To leading order around flat space, the dCS operator is schematically (suppressing indices and $\mathcal{O}(1)$ factors) the dimension 7 operator
\begin{equation}
    \mathcal{L}_\text{dCS}\sim {\alpha\over\mpl}\,\phi\,\partial\partial h_c\,\partial\partial h_c
\end{equation}
where $h_c\equiv h/\mpl$ is the canonically normalized graviton field.
The 2-to-2 scattering  amplitude is then estimated as
\begin{equation}
\mathcal{A}_{2\to 2}\sim {\alpha^2\over \mpl^2} E^6.
\end{equation}
Therefore, to obey unitarity $\mathcal{A}_{2\to 2}\lesssim 1$, the cutoff for the theory is\footnote{For a brief introduction of partial wave unitarity constraints, please see \cite{Schwartz:2014sze}.}
\begin{equation}
    \Lambda_\text{unitarity}\sim \left(\mpl\over|\alpha|\right)^{1/3}.
\label{unitarity}\end{equation}  
Note that this is parametrically larger than the inverse length scale $1/\sqrt{|\alpha|}$ that appears in the Lagrangian. We shall soon see that constraints from causality are much more severe than this. These points were already noted in Ref.~\cite{Serra:2022pzl}.

\subsection{Perturbations and Dispersion Relations}\label{subsec: Perturbed Equations of Motion}

In order to examine signal speeds around a background, we now perturb the equations of motion up to first order in perturbations \eqref{eq: metric eom} and \eqref{eq: scalar eom} via \eqref{eq: metric perturbation def} and \eqref{eq: phi perturbation def}. 
We  impose the transverse-traceless (TT) gauge, namely that $\overline{g}^{ab}h_{ab} \equiv h^a{}_a = 0$, and $\overline{\nabla}_a h^{ab} = 0$. In doing so, we get a large number of terms, most of which vanish for our particular purposes when looking at backgrounds such that $\overline{R}_{ab}=0$ which allows us to derive a dispersion relation. The full perturbed equations of motion can be found in the corresponding \textit{Mathematica} notebook, while the simplified version can be found in \eqref{eq: appendix perturbed scalar eom} and \eqref{eq: appendix perturbed metric eom}.

For the gravitational wave backgrounds studied here, the Pontryagin density vanishes $^*\overline{R}\,\overline{R}=0$. 
And (as discussed above) we set the potential to vanish $V=0$.
This implies a valid solution is $\overline{\phi} = 0$ as well since the source term in the background equation of motion for the scalar field vanishes. 
So the background has the property
\begin{equation}
^*\overline{R}\,\overline{R}=0,\,\,\,\,\,\,
\overline{R}^{ab}=0,\,\,\,\,\,\,\overline{\phi}=0.
\end{equation}
After performing this simplification, we have the following two equations of motion for the perturbations
\begin{align}
    \overline{\nabla}^c\overline{\nabla}_ch^{ab} & = {2\alpha\over\mpl} {}^*\overline{R}^b{}_c{}^a{}_d\overline{\nabla}^c\overline{\nabla}^d\delta\phi\label{eq: metric eom for R_{ab} = 0}\\
    \overline{\nabla}_a\overline{\nabla}^a\delta\phi & = -\alpha\,\mpl\left({}^*\overline{R}_{bcde}\overline{R}_a{}^{cde}h^{ab} + {}^*\overline{R}_{adbc}\overline{\nabla}^d\overline{\nabla}^ch_{ab}\right)\label{eq: scalar eom for R_{ab}=0}.
\end{align}

We can now Fourier transform and solve for a (non-linear) dispersion relation. Since there are 3 degrees of freedom (2 modes of the graviton plus 1 mode of the scalar), we must have 3 eigen-modes of the system. 1 eigen-mode is the following: we set $\delta\phi=0$, the first equation is then just the standard wave-equation in vacuum for the metric, giving $k_a k^a=0$; i.e., the standard dispersion relation. In order for this to be consistent, the second equation demands that $h_{ab}$ obey a special rule; this will only be satisfied by some special combination of the 2 graviton modes. This single eigen-mode is not interesting from the point of view of causality as its dispersion relation is standard.

The other 2 eigen-modes are much more interesting and will be our focus. For these modes, we have $\delta\phi\neq0$ (as well as $h_{ab}\neq0$).
The details of this analysis can be found in Appendix \ref{sec: appendix for perturbed eom and consistency}. We find the following dispersion relation for a wave packet sent through the background
\begin{equation}
    (k_ak^a)^2 = 2\,\szeta^2 \,{}^*\overline{R}_{adbc}{}^*\overline{R}^b{}_e{}^a{}_fk^dk^ck^ek^f
    .
    \label{eq: dispersion relation in paper}
\end{equation}
Here \eqref{eq: dispersion relation in paper} matches a result given in Ref.~\cite{Garfinkle:2010zx}\footnote{To match their result, recall that the Riemann tensor is exactly the Weyl tensor when $\overline{R}_{ab} = 0$.}.

\section{Time Delay on a Gravitational Wave Background}\label{sec: Shapiro Time Delay on a Gravitational Wave Background}

In this section, we put our dispersion relation \eqref{eq: dispersion relation in paper} that is already on a simplified background $\overline{R}_{ab} = 0$, into a specific metric that obeys this constraint, namely a gravitational wave background metric. We then compute the frequency, and then also find the corresponding velocity. This allows us to find the time delay by integrating over the velocity. We then show that in dCS gravity on a gravitational wave background, there exist superluminal modes that are not saved by the  Shapiro time delay that we compute up to second order. Then, we derive an inequality that dCS must obey in order to not produce superluminal signals. To end the section, we discuss implications for observations.

\subsection{Frequency and Velocity Relation}\label{subsec: Frequency and Velocity Relation}

Consider a background that is a gravitational wave. We write it as 
\begin{equation}
    \overline{g}_{ab} = \eta_{ab} + \h_{ab}.
\end{equation}
For a plane wave in the $z$-direction, we can write this as
\begin{equation}
    \overline{g}_{ab} = 
    \begin{pmatrix}
        -1&0&0&\\
        0&1+\h_+(t-z)&\h_\times(t-z)&0\\
        0&\h_\times(t-z) & 1-\h_+(t-z) & 0\\
        0&0&0&1
    \end{pmatrix}\label{eq: background metric matrix form}
\end{equation}
where the background gravitational polarizations are $\h_+$ (plus) and $\h_\times$ (cross). 
We specify the covariant momentum components as $k_a = (-\omega_p,k_x,k_y,k_z)$. We refer to $\omega_p$ as the characteristic frequency of the perturbation. The dispersion relation \eqref{eq: dispersion relation in paper}  has $2$ (positive) solutions of $\omega_{p}$. After some considerable work, these are found to be
\begin{align}
    \omega_{p}^2 = \left(1 - \h_+\right)k_x^2 - 2 \h_\times k_xk_y + (1+ \h_+)k_y^2 + k_z^2\pm \szeta\,(\omega_p-k_z)^2\sqrt{(\h_\times'')^2 + (\h_+'')^2}\label{eq: frequency solution part 1}.
\end{align}
Taking the square root and working to first order while using $|k|^2=|{\bf k}|^2 = k_x^2+k_y^2+k_z^2$ gives
\begin{equation}
    \omega_{\pm} = |k| + \frac{1}{2|k|}\left(\h_+(k_y^2-k_x^2)-2\h_\times k_xk_y\pm\szeta\,(|k|-k_z)^2\sqrt{(\h_\times'')^2+(\h_+'')^2}\right)
    \label{eq: final answer for frequency}.
\end{equation}

Let us now compute the associated velocity. The phase and group speeds are 
\begin{equation}
v_\text{phase}={\omega_p\over|k|},\,\,\,\,
\,\,\,\,\,\,\,\,
v_\text{group}=\left|{\partial\omega_p\over\partial{\bf k}}\right|.
\end{equation}
After taking these partial derivatives, combining, and again working to first order, we find that the phase and group speeds are the same. They are
\begin{equation}
    v_\pm =1 - \frac{1}{2}\h_+ + \frac{1}{2|k|^2}\left(\h_+k_z^2+2\h_+k_y^2-2\h_\times k_xk_y\right) \pm \frac{\szeta\,(|k|-k_z)^2}{2|k|^2}\sqrt{(\h_\times'')^2+(\h_+'')^2}\label{eq: velocity relation}.
\end{equation}

\subsection{Total Time Delay}\label{subsec: Total Time Delay}

\subsubsection{Gaussian Wave Packets \& Shapiro at First Order}\label{subsubsec: gaussian wave packets and first order shapiro}

\begin{figure}[t]
    \centering
    \includegraphics[width=1.0\linewidth]{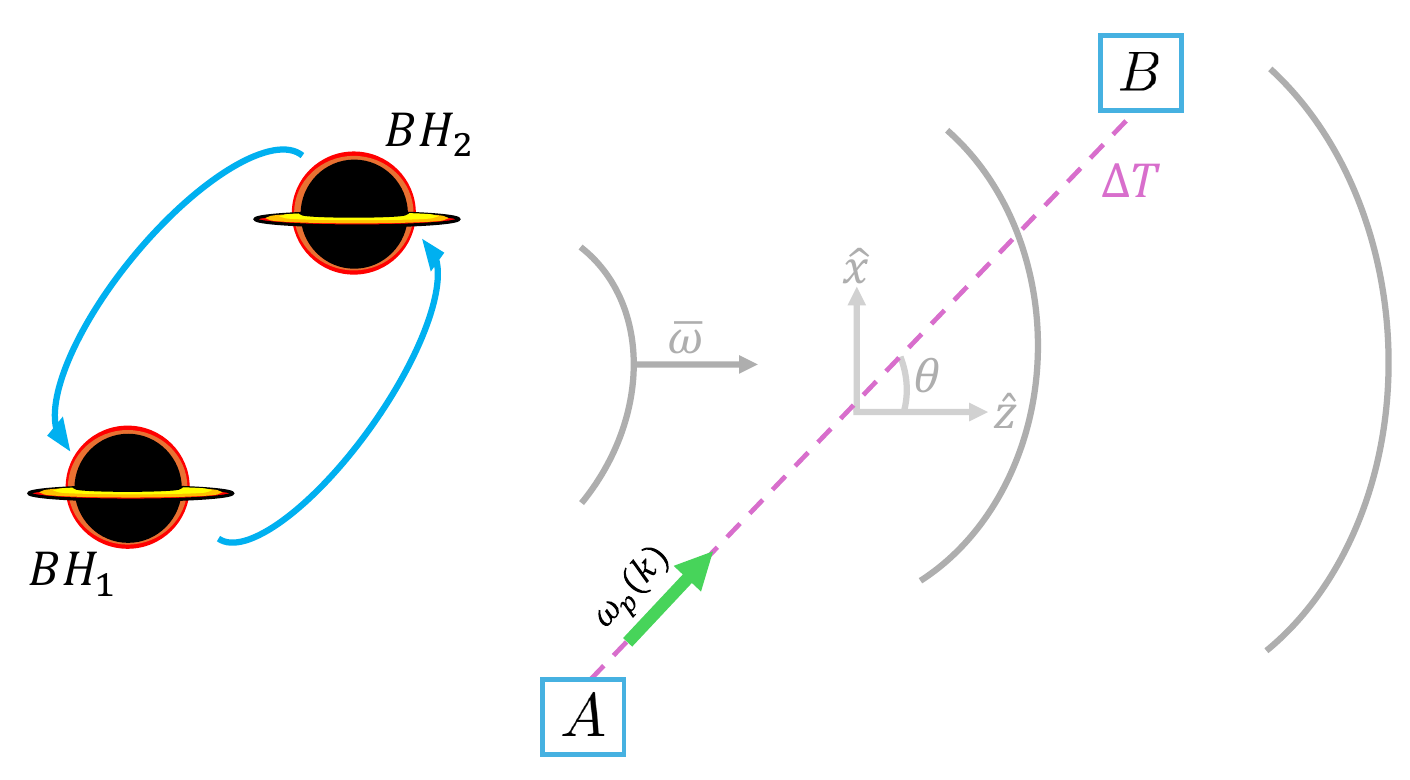}
    \caption{Time delay across a gravitational wave generated by an orbiting system (black holes).}
    \label{fig:time delay bh image}
\end{figure}

To compute the Shapiro time delay (or simply the time delay), we consider the following setup given in figure \ref{fig:time delay bh image}.

We picture a scenario where the production of gravitational waves is facilitated by the spiraling of two masses $M_1$ and $M_2$ (for example, one can consider a binary black hole system of the sort LIGO/Virgo is observing of several solar masses and size 10's of km). 
Then we imagine a signal (wave packet) is sent from A to B, acting as our probe for measuring the time delay. We will eventually take $(A,B)\rightarrow (-\infty,\infty)$ and compare starting an ending regions that are asymptotically Minkowski. The coordinate map we superimpose on the gravitational wave can be seen in figure \ref{fig:time delay bh image}, which implies the following unit vectors
\begin{equation}
\hat{k}_x = \sin\theta\cos\varphi,
\,\,\,\,\,\hat{k}_y = \sin\theta\sin\varphi, 
\,\,\,\,\,\hat{k}_z = \cos\theta.
\end{equation}
These obey $\sum_{i=1}^3\hat{k}_i^2 = 1$. From the normal vectors, we can construct projections $P_+(\hat{k}) = \hat{k}_x^2 - \hat{k}_y^2 = \sin^2\theta\cos(2\varphi)$ and $P_\times(\hat{k}) = 2\hat{k}_x\hat{k}_y =  \sin^2\theta\sin(2\varphi)$. This allows us to write \eqref{eq: velocity relation} in a nice form as
\begin{equation}
    v_\pm = 1-\frac{1}{2}\left(\h_+P_+ + \h_\times P_\times\right)\pm\frac{\szeta}{2}(1-\cos\theta)^2\sqrt{(\h_+'')^2 + (\h_\times'')^2}.\label{eq: nice velocity relation}
\end{equation}
The time delay from A to B can be written as an integral over the path and approximated as
\begin{equation}
    \Delta T_\pm = \int_{A}^Bd\ell \left(\frac{1}{v}-1\right) \simeq -\int_A^Bdt\,(\delta v)\label{eq: time delay integral},
\end{equation}
where $\delta v=v-1$ is the perturbation in speed from unity.
We plug-in \eqref{eq: nice velocity relation} giving us the following form
\begin{align}
    \Delta T_\pm &\simeq \frac{1}{2}\int_A^Bdt(\h_+P_+ + \h_\times P_\times) \mp \frac{\szeta}{2}(1-\cos\theta)^2\int_A^Bdt \sqrt{(\h_+'')^2 + (\h_\times'')^2}\label{eq: total shapiro time delay}.
\end{align}
The first term is from GR (related to Shapiro time delay), while the second term is from dCS.

For infinite plane waves, these integrals diverge. Instead, let us consider a ``nearly'' plane wave that has a Gaussian profile. We assume it is dominated by a single frequency $\overline{\omega}$ and has a width $\sigma$ that is large, i.e.,
\begin{equation}
\overline{\omega}\,\sigma\gg 1.
\end{equation}
To make the above dCS integral simple, we shall consider the waves $\h_+$ and $\h_\times$ to be out of phase with the same amplitude, as
\begin{eqnarray}
\h_+(u) = H\,E(u)\cos(\overline{\omega}u + \phi_0)\:\:\:\text{and}\:\:\: 
\h_\times(u) = H\, E(u)\sin(\overline{\omega}u + \phi_0)\label{eq: general wave form packet}
\end{eqnarray}
where $H$ is the amplitude of the wave packet, the spacetime argument is
\begin{equation}
u\equiv t-z=t(1-\cos\theta),
\end{equation}
$\phi_0$ is a generic phase shift, and $E(u)$ denotes a Gaussian function. This choice makes the $\sqrt{(\h_+'')^2+(\h_\times'')^2}$ factor very simple because we obtain a factor of $\sqrt{\cos^2(\overline\omega u)+\sin^2(\overline\omega u)}=1$ times a Gaussian.

For a wave of finite width ($\sigma$) in the forward $z$-direction, but infinite width in the transverse direction, we have 
\begin{equation}
    E(u) = \exp(-u^2/2\sigma^2).
\end{equation} 
For generic angles, this is integrable.
We could also consider a  plane wave that is infinite in the forward direction but finite ($\sigma$) in the transverse $xy$-directions.
This has a Gaussian of the form $E = \exp(-(x^2+y^2)/2\sigma^2)$ where $\sigma$ is now the transverse width of the wave. In our coordinate system, this wave can be written as $x^2 + y^2 = t^2\sin^2\theta = u^2\cot^2(\theta/2)$, allowing us to re-write the Gaussian as 
\begin{equation}
    E(u) = \exp(-u^2/2\sigma^2_\text{eff}),\,\,\,\,\,
    \mbox{with}\,\,\,\,\,\,\sigma_\text{eff}\equiv \sigma\, |\tan(\theta/2)|.
\end{equation} 
Therefore, all of our results in analyzing the case in which the wave is modulated in the forwards direction can be carried over to the case in which the wave is modulated in the transverse directions by the replacement $\sigma\to \sigma_\text{eff}$. 

In the wide Gaussian limit ($\overline\omega\,\sigma\gg 1$), when plugging the wave form \eqref{eq: general wave form packet} into \eqref{eq: total shapiro time delay}, we find the following time delay (taking $H>0$)
\begin{equation}
\Delta T_\pm=\Delta T_\text{GR}+\Delta T_\text{dCS}
\end{equation}
where the GR and dCS pieces are
\begin{align}
    \Delta T_\text{GR}&\approx H\frac{\sqrt{2\pi}\,\sigma}{2(1-\cos\theta)}e^{-(\overline{\omega}\,\sigma)^2/2}\left(P_+\cos\phi_0 + P_\times\sin\phi_0\right)\\
     \Delta T_{\text{dCS}\pm}&\approx \mp H\,\szeta\,\sqrt{\pi}\,(1-\cos\theta)\,\overline{\omega}^2\sigma
    \label{eq: total time delay computed in wide gaussian limit}
\end{align}
up to $\mathcal{O}((\overline{\omega}\sigma)^{-1})$ corrections. 

We see that in the wide Gaussian limit, the GR term here is exponentially suppressed, while the dCS term is not. Hence we have
\begin{equation}
    \Delta T_\pm\approx\Delta T_{\text{dCS}\pm}.
\end{equation}
Since one of the modes has a negative value for $\Delta T_\text{dCS}$, we can have time {\it advance} $\Delta T<0$, i.e., superluminality. To be sure this is measurable, we impose that the advance is larger than our resolution limit. This can be taken to be the inverse of frequency of the perturbation, i.e., 
\begin{equation}
    |\Delta T|>\delta T_\text{res}\sim{1\over\omega_p}.
\label{measure}\end{equation}
But this measurability condition is always achievable by simply making the width $\sigma$ arbitrarily large in Eq.~(\ref{eq: total time delay computed in wide gaussian limit}). 

This appears to indicate we have superluminality for any nonzero value of the dCS coupling $\alpha$. However, this conclusion is too hasty; the above follows from the GR term (Shapiro) being exponentially small, and while it is so at first order, it will not be so at second order, as we now examine.

\subsubsection{Shapiro at Second Order}\label{subsubsec: shapiro at second order}
To compute the GR (Shapiro) time delay at second order in metric perturbations, we specialize our wave packet to another simple case in which the two polarizations are in phase, as  
\begin{equation}
    \h_\times = \h_+ = H\,E(u)\cos(\overline{\omega}u)
\end{equation} 
(again with $E(u)=e^{-u^2/2\sigma^2}$ a Gaussian and $u=t-z$).
This simplifies the computation at second order and matches the form studied in Ref.~\cite{Misyura:2025lcr} which we restate for clarity\footnote{In the linked \textit{Mathematica} notebook, we show explicitly that the metric is a solution to $\overline{R}_{ab} = 0 + \mathcal{O}(H^3)$.}
\begin{align}
    \g_{00} & = -1 + \frac{\HH^2}{8}\left(6t^2\overline{\omega}^2-4t\overline{\omega}^2z^2 + 2\overline{\omega}^2z^2 + 6\sin(2\overline{\omega}(t-z))t\overline{\omega} - 3\cos(2\overline{\omega}(t-z))+3\right)\nonumber\\
    \g_{03} & = \frac{\HH^2}{8}\left(-3-4t^2\overline{\omega}^2+8t\overline{\omega}^2z - 6\sin(2\overline{\omega}(t-z))t\overline{\omega}+3\cos(2\overline{\omega}(t-z))\right)\nonumber\\
    \g_{11} & = 1  + \HH\cos(\overline{\omega}(t-z))+\frac{\HH^2}{2}\cos(2\overline{\omega}(t-z))  - \frac{3\HH^2}{4}\sin^2(\overline{\omega}(t-z)) - \frac{\HH^2}{4}\overline{\omega}^2(t-z)^2\nonumber\\
    \g_{22} & = 1- \HH\cos(\overline{\omega}(t-z))-\frac{\HH^2}{2}\cos(2\overline{\omega}(t-z))  - \frac{3\HH^2}{4}\sin^2(\overline{\omega}(t-z)) - \frac{\HH^2}{4}\overline{\omega}^2(t-z)^2\nonumber\\
    \g_{12} & =  \HH\cos(\overline{\omega}(t-z))+\frac{\HH^2}{2}\cos(2\overline{\omega}(t-z)) \nonumber\\
    \g_{33} & = 1 + \frac{\HH^2}{8}\left(6t^2\overline{\omega}^2-4t\overline{\omega}^2z + 2\overline{\omega}^2z^2 + 6\sin(2\overline{\omega}(t-z))t\overline{\omega} - 3\cos(2\overline{\omega}(t-z)) + 3\right)\label{eq: second order metric}.
\end{align}
where we have written the amplitude as $\HH=H\,E(u)$. This form is only exact (at this order) for constant $E(u)$, but is approximately correct for a slowly varying modulation in $E(u)$.

Recall that only the GR term was exponentially suppressed at first order. So we only need to re-compute this piece at second order. This is the correction in speed from the standard $k_ak^a=g^{ab}k_ak_b=0$ dispersion relation. Plugging the metric \eqref{eq: second order
metric} into this, and working to second order in $H$ with $h(u) = H\cos(\overline{\omega}u)e^{-u^2/2\sigma^2}$ as well as in the wide Gaussian limit, $\overline{\omega}\,\sigma\gg1$, we find the following contribution to the GR (Shapiro) time delay
\begin{equation}
    \Delta T^{(2)}_\text{GR}\approx H^2\sqrt{\pi}\,\overline{\omega}^2\sigma^3g(\theta)\label{eq: second order shapiro time delay}
\end{equation}
where the function $g(\theta)$ is
\begin{align}
    g(\theta) & = \frac{1}{128}\left(2\hat{k}_x^2 + 2\hat{k}_y^2 + 7\hat{k}_z^2 - 4\hat{k}_z + 4\cos\theta(\hat{k}_x^2 + \hat{k}_y^2 + 2\hat{k}_z) + \hat{k}_z^2\cos(2\theta)\right)\csc^6(\theta/2)\nonumber\\
    & = \frac{1}{512}\left(35-12\cos\theta+28\cos(2\theta) - 4\cos(3\theta) + \cos(4\theta)\right)\csc^6(\theta/2)\label{eq: shapiro second order g-theta function}
\end{align}
where in the second line we substituted in the values for $\hat{k}_x$, $\hat{k}_y$, and $\hat{k}_z$. 

Let us comment on a comparison between the group and phase speeds here. We can expand the frequency as $\omega_\text{GR} = |k| + \delta\omega_\text{GR}^{(1)} + \delta\omega_\text{GR}^{(2)}$ where $\delta \omega_\text{GR}^{(1)}$ denotes the leading order deviation from the Minkowski dispersion relation in equation \eqref{eq: final answer for frequency} as
\begin{equation}
    \delta\omega_\text{GR}^{(1)}
    =\frac{1}{2|k|}\left(\h_+(k_y^2-k_x^2)-2\h_\times k_xk_y\right).
\end{equation}
Then, working to second order, one can show that the group speed and the phase speeds are related by
\begin{equation}
    v_\text{group}=v_\text{phase}+{1\over2}|k|^2\left|{\partial(\omega_\text{GR}^{(1)}/|k|)\over\partial{\bf k}}\right|^2.
\label{groupphase}\end{equation}
The second term here is on the order $H^2$. Then when carrying out the integral with the Gaussian to obtain the time delay, this second term gives a contribution of order $H^2\,\sigma$. This is subdominant to the above term that is of the order $H^2\overline{\omega}^2\sigma^3$. Hence phase and group speeds again give the same result to the order we are working. (Similarly, we do not need to include the $(\delta v)^2$ correction to the time delay to this order.) 

We note that the result in equation \eqref{eq: second order shapiro time delay} of $\Delta T_\text{GR}\sim H^2\overline\omega^2\sigma^3$ is reasonable: for the standard Shapiro time delay near a static source, recall that it is $\Delta T_\text{GR}\sim G\,M$ (up to some log-factors). If one applies this reasoning to a gravitational wave, one can estimate $M\sim\rho_\text{GW}\,V$, where $\rho_\text{GW}\sim(\h')^2/G\sim\overline\omega^2H^2/G$ is the energy density and $V\sim\sigma^3$ is the volume over which it is appreciable, thus giving the estimate $\Delta T_\text{GR}\sim H^2\overline\omega^2\sigma^3$. The precise computation also gives the $\sqrt{\pi}\,g(\theta)$ factor above.


For the dCS contribution, we only need to work to first order. This is essentially identical to that computed in the previous subsection, except the 2 polarizations are now in phase instead of out of phase. This means from $\sqrt{(\h_+'')^2+(\h_\times'')^2}$, we have a factor of $\sqrt{\cos^2(\overline\omega u)+\cos^2(\overline\omega u)}=\sqrt{2}|\cos(\overline{\omega}u)|$. Carrying out this integral gives a slight reduction in the result by a factor
\begin{equation}
    \factor={2\sqrt{2}\over\pi}\approx0.9
\end{equation}
The dCS result is then
\begin{equation}
    \Delta T_{\text{dCS}\pm}\approx \mp H\,\szeta\,\factor\,\sqrt{\pi}\,\overline{\omega}^2\sigma\,\widetilde{g}(\theta)
\end{equation}
where $\widetilde{g}(\theta) = (1-\cos\theta)$.

Since the GR term is negligible at first order, we combine just the second order GR term and the first order dCS term to give the total time delay as
\footnote{Note that this is qualitatively different from some previous works when considering dispersion relations and time delay. For example, in \cite{Gruzinov:2006ie} they considered an EFT of GR that contained quartic Riemann curvature operators. Then, by enforcing that the correction to standard propagation is purely positive, they were able to obtain the condition that the coefficients are positive.
 However, their analysis drops the corresponding Shapiro term, which is only valid in a wide Gaussian limit that we considered here. It would therefore be interesting to see what happens if their analysis is redone but by considering a specific wave packet and going to second order in GR.}
\begin{align}
    \Delta T_\pm &\approx \Delta T_\text{GR}^{(2)}+\Delta T_{\text{dCS}\pm}\\
&\approx    H^2\sqrt{\pi}\,\overline\omega^2\,\sigma^3g(\theta)
\mp H\,\szeta\,\factor\sqrt{\pi}\,\overline{\omega}^2\,\sigma \,\widetilde{g}(\theta).
\label{eq: total time delay with second order term}
\end{align} 

\subsection{Causality Constraint on Coupling}

Again by focusing on the mode with the minus sign in the dCS term, there is a possibility of time advance with $\Delta T<0$. For large gravitational wave amplitude $H$, the GR (second order Shapiro) term dominates and we have $\Delta T>0$, while for small $H$, the dCS term dominates and we have $\Delta T<0$ (for one of the modes). However, for this to be measurable, we again impose that the time advance is larger than the resolution $|\Delta T|>\delta T_\text{res}\sim1/\omega_p$, as given earlier in Eq.~(\ref{measure}).
So we cannot simply make the amplitude $H$ arbitrarily small, as this will render $|\Delta T|$ so small the effect is not measurable.

The most dangerous situation is therefore when $H$ is just small enough for the dCS term to be an $\mathcal{O}(1)$ factor larger than the GR term. The most negative value of $\Delta T$ occurs when the dCS is -2 times the GR term
\begin{equation}
    |\Delta T_{\text{dCS}}|=2\Delta T_\text{GR}^{(2)}
\end{equation}
Solving for this special amplitude (we call it $H^*$) gives
\begin{equation}
    H^*={\szeta\,\factor\,\widetilde{g}(\theta)\over2\,\sigma^2\,g(\theta)}.
\end{equation}
The corresponding (most negative) value of $\Delta T$ is
\begin{equation}
    \Delta T^*=-{\szeta^2\factor^2\sqrt{\pi}\,\overline\omega^2\,\widetilde{g}(\theta)^2\over4\,\sigma\,g(\theta)}.
\end{equation}
In order to avoid measurable superluminality, we now impose $|\Delta T^*|\lesssim 1/\omega_p$, giving a bound on the dCS coupling of
\begin{equation}
       |\alpha|\lesssim  {s(\theta)\over\overline\omega^2},\label{eq: alpha bound in terms of s(theta)}
\end{equation}
where we defined the prefactor $s(\theta)$ as
\begin{equation}
s(\theta)\equiv
\frac{2\sqrt{\chi}\,\sqrt{g(\theta)}}{\factor\,\pi^{1/4}\sqrt{\gamma}\,\widetilde{g},(\theta)}
\label{sfactor}\end{equation}
and introduced a pair of dimensionless parameters
\begin{equation}
    \chi\equiv\overline\omega\,\sigma\,\,\,\,\text{and}\,\,\,\,
    \gamma\equiv{\omega_p\over\overline\omega}.
\end{equation}
By pushing $\overline\omega$ to its highest value allowed by causality in \eqref{eq: alpha bound in terms of s(theta)}, which we denote $\Lamc$, the causality cutoff is
\begin{equation}
|\alpha|={s(\theta)\over\Lamc^2}
\implies
\Lamc=\sqrt{\frac{s(\theta)}{|\alpha|}}
\label{cbd}\end{equation}
From Eq.~(\ref{cbd}) we see that the causality cutoff is on the order of the length scale in the Lagrangian $\Lamc\sim1/\sqrt{|\alpha|}$, unlike the  cutoff from unitarity in Eq.~(\ref{unitarity}) of $\Lambda_\text{unitarity}\sim(\mpl/|\alpha|)^{1/3}$, which is parametrically larger. So the causality bound is more constraining.

\begin{figure}[t]
    \centering
    \includegraphics[width=0.7\linewidth]{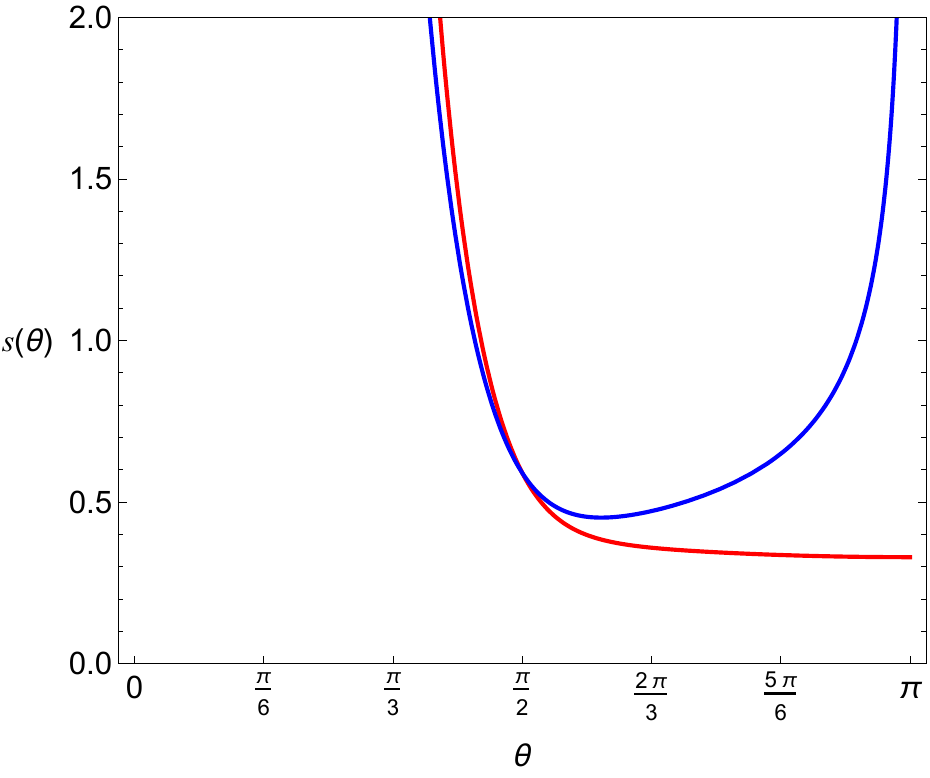}
    \caption{Plot of the function $s(\theta)$ from Eq.~(\ref{sfactor}) versus angle $\theta$. Here we have set $\chi=\gamma$ for concreteness.
    The {\color{red}{red}} curve is for a Gaussian with modulation in the forwards direction, while the {\color{blue}{blue}} curve is for a Gaussian with modulation in the transverse direction.}
    \label{sfig}
\end{figure}

The precise bound depends on the value of $s(\theta)$. We have provided a plot of this in Figure \ref{sfig}. 
The parameter $\chi=\overline\omega\,\sigma$ must be large, as discussed earlier, to justify the plane wave approximation. And the parameter $\gamma=\omega_p/\overline\omega$ must also be large in order to treat the wave packet (the signal) as small ($L_\text{wp}\sim 1/\omega_p$) compared to the characteristic size of variation the background ($L_\text{bgd}\sim1/\overline\omega$). For concreteness, in Figure \ref{sfig} we have taken $\chi=\gamma$. Otherwise, if they are not equal, the plot is rescaled by a factor of $\sqrt{\chi/\gamma}$.  Finally, we  must choose to either consider the wave with profile modulated in the forwards direction (red curve) or modulated in the transverse direction (blue curve). For the latter, recall that we replace $\sigma\to\sigma_\text{eff}=\sigma\,|\tan(\theta/2)|$. Since $\sigma$ appears in $\chi$, this means that there is an additional factor of $\sqrt{|\tan(\theta/2)|}$ in this second case relative to the first.

From the figure, we see that the minimum value of $s$ is
\begin{equation}
    s_\text{min}\approx 0.33\sqrt{\chi\over\gamma}\,\,\,\,(\mbox{forwards case}),\,\,\,\,\,\,\,
    s_\text{min}\approx 0.45\sqrt{\chi\over\gamma}\,\,\,\,(\mbox{transverse case}).
\end{equation}
The minimum value provides the sharpest causality bound; so this comes from the forwards case. For concreteness, let us take $\chi\sim\gamma\gg1$. This gives our final estimate for the causality bound as
\begin{equation}
|\alpha|={s_0\over\Lamc^2}
\implies
\Lamc={\sqrt{s_0}\over\sqrt{|\alpha|}},
\,\,\,\,\,\,
\mbox{with}\,\,\,\,s_0\sim 0.3
\label{c03}\end{equation}

We note that this bound is roughly compatible with a known bound on dCS in the literature from considering a wave packet moving near a static massive object \cite{Serra:2022pzl}. This static source analysis gives a roughly similar result for $\Lamc$, albeit it is enhanced by a log factor and not suppressed by the $\gamma=\omega_p/\bar{\omega}$ factor. So our result here from the dynamical gravitational wave source is moderately sharper.


\subsubsection{Limit of Applicability}

One might try to push this even further by taking $\gamma$ even much larger than $\chi$, thus lowering $s_0$ even further. Formally, the true signal speed is given by sending the frequency of the wave packet $\omega_p\to\infty$ \cite{Brillouin1960}.
Then the phase speed gives the wavefront speed of a localized configuration.
This sends $\gamma\to\infty$ and $s_\text{min}\to0$. So this would push $\alpha\to0$ to avoid superluminality. This would be a very strong conclusion. 

However, at asymptotically large $\omega_p$, it is unclear if one can trust the effective theory with such a large value of the frequency of the wave packet. Moreover, the above theory has a kind of UV completion as we discuss in the next section.
In any case, we take the above as a conservative estimate. (One might consider $\chi\sim\gamma\sim 10$ to justify both requirements.)

\subsection{Implication for Observations}

Let us check what the implication of this bound is for the dCS term to impact observations. Suppose we are considering systems in the strong gravity regime, such as black holes or neutron stars. Let us denote the length scale we are probing as $L$ (with $L$ the horizon size of the black hole or radius of neutron star). Then the Riemann tensor can be estimated as $R\sim 1/L^2$. 

Let us now estimate the size of the terms in the action. The Einstein-Hilbert and dynamical Chern-Simons terms are
\begin{align}
&\mathcal{L}_\text{EH}    =\mpl^2\,R\sim{\mpl^2\over L^2}\\
&\mathcal{L}_\text{dCS}={\alpha\,\mpl\over4}\phi\,^*R\,R\sim{\alpha\,\mpl\,\phi\over L^4}.
\end{align}
Here we are assuming the object significantly breaks parity, such as a rapidly rotating Kerr black hole, for which the Pontryagin density can indeed be estimated as $^*R\,R\sim 1/L^4$; otherwise there is an additional suppression for nearly parity even states. 
The equation of motion for $\phi$ allows us to estimate it as 
\begin{equation}
    \Box\phi=-{\alpha\,\mpl\over4}\,^*R\,R\implies {\phi\over L^2}\sim{\alpha\,\mpl\over L^4}.
\label{phiEst}\end{equation}
Inserting this into $\mathcal{L}_\text{dCS}$ gives an estimate for the correction to GR provided by dCS as
\begin{equation}
    {\mathcal{L}_\text{dCS}\over\mathcal{L}_\text{GR}}
    \sim {\alpha^2\over L^4}\equiv\delta\implies |\alpha|=\sqrt{\delta}\,L^2.
\label{alphadL}\end{equation}
  We have denoted this as $\delta$, which is an estimate for the fractional correction to the dynamics provided by dCS. So, for example, if $\delta=0.1$, then dCS is correcting the dynamics in the strong gravity regime (of parity breaking states) at the $\sim 10\%$ level. For LIGO/Virgo observations of merging black holes, we may be sensitive to $\delta$ of a few percent, but not much smaller at this stage.

  Inserting the expression for $\alpha$ in terms of the causality scale (\ref{c03}) into this and expressing the result in terms of  $\delta$ gives
  \begin{equation}
      \delta=s_0^2\left(1\over L\,\Lamc\right)^4.\label{eq: delta causality}
  \end{equation}
Now let us see why this is important. 
We need that the scale that we are probing $L$ must be larger than the scale $\Lamc^{-1}$ or else there would be causality breakdown, 
\begin{equation}
L>\Lamc^{-1}.  
\end{equation}
So the term in brackets in \eqref{eq: delta causality} must be small, and it is raised to the power of 4, therefore rendering it very small. Also, we know from above that $s_0\sim0.3$, and squaring it gives another small factor. 
So overall this implies $\delta$ must be very small. 

We can go a little further and consider the impact of new physics that must kick in to restore causality; an explicit example will be given in the next section. Let us suppose that there is new physics associated with new degrees of freedom of mass $m_\psi$ that we integrate to generate the dCS term. We can rewrite the above expression for $\delta$ in terms of this as
\begin{equation}
    \delta=s_0^2\left(1\over L\,m_\psi\right)^4\left(m_\psi\over\Lamc\right)^4
\label{bddeltaLm}\end{equation}
where we have simply multiplied and divided by $m_\psi^4$. The reason this is useful to do is that we have the following pair of inequalities: (i) the scale we are probing $L$ must be larger than the Compton wavelength of the new degrees of freedom $m_\psi^{-1}$ or else we could not use the dCS effective theory,
\begin{equation}
    L>m_\psi^{-1}.
\label{LmBd}\end{equation}
And (ii), the new physics itself must enter before we have causality breakdown
\begin{equation}
m_\psi<\Lamc.
\label{mLamBd}\end{equation}
Then from Eq.~(\ref{LmBd}) the first term in brackets in (\ref{bddeltaLm}) must be small, and from Eq.~(\ref{mLamBd}) the second term in brackets must also be small. Then, since these terms are raised to the power of 4 and combined with $s_0\sim0.3$, this really emphasizes that $\delta$ must be very small.
This implies that the size of the effects of the dCS term (which are estimated as $\delta$) must be quite suppressed even for strong gravity systems. 
For LIGO/Virgo, we are perhaps only sensitive to $\delta$ of a few percent, as mentioned above, which seems too large compared to this theoretical bound.
Also note that for systems in the weak gravity regime, we are even much less sensitive to this operator.

\section{UV Completion}\label{sec: Constraints from Unitarity, and UV Completions}


In this section, we examine a UV completion of dCS. We combine this with the gravitational species bound to find another very sharp inequality. We then use the UV completion to mention other operators and their consequences on the UV completion and our inequalities.

\subsection{UV Completion from Fermions}
 Consider a set of $N$ Dirac fermions $\psi_n$, chirally coupled to a scalar $\phi$. The action is
 \begin{equation}
    S = \int d^4x\sqrt{-g}\left[\mpl^2\,R-{1\over2}g^{ab}\nabla_a\phi\nabla_b\phi+\sum_n\overline\psi_n(i\,\gamma^a\nabla_a - m_n +i\,g_n\,\phi\,\gamma_5)\psi_n\right]
    \label{eq: uv completion action of dCS}
\end{equation}
where $m_n$ is the mass of each species and $g_n$ is the (pseudo) Yukawa coupling. 

We note that this action breaks the shift symmetry in $\phi$. However, we can view this action as just the leading part of a larger theory in which $\phi$ is the Goldstone boson of a spontaneously broken global U(1). To do so, we can introduce a complex scalar field $\Phi$, with coupling to fermions 
\begin{equation}
   S_\Phi = \int d^4x\sqrt{-g}\left[\mpl^2\,R
       -{1\over2}|\partial\Phi|^2+
       \sum_n
       \overline{\psi}_n
       \big{(}i\,\gamma^a\nabla_a-
       g_n\,\overline{\psi}_n\,(\Phi^*\, P_R+\Phi \,P_L)\,\psi_n\big{)}
   -V(|\Phi|) \right]\!.
   \label{eq: Phi uv completion action of dCS}
\end{equation}
This theory has the global U(1) symmetry
\begin{equation}
    \Phi\to e^{i\theta_0}\,\Phi,\,\,\,\,\,\,\,\,
    \psi_n\to e^{i\theta_0\gamma_5/2}\,\psi_n.
\end{equation}
One endows $\Phi$ with a potential that spontaneously breaks the symmetry
\begin{equation}
    V(|\Phi|)\propto\left(|\Phi|^2-f^2\right)^2.
\end{equation}
Then by expanding around the vacuum
\begin{eqnarray}
    \Phi=f\,e^{i\phi/f},
\end{eqnarray}
where $\phi$ is the Goldstone boson,
and Taylor expanding to linear order in $\phi$, we obtain the above action in (\ref{eq: uv completion action of dCS}) with fermion masses given by $m_n=g_n\,f$.

It is known that integrating out the fermions generates the dynamical Chern-Simons term. This is also closely related to the chiral anomaly. There are two 1-loop diagrams, as given here
\begin{equation}
\TriangleAHHLoop[0.5]\,\,\,\,\,\,\,\,\,\,\,\,\,\,
\CircleAHHLoop[0.5]
\label{eq: feynman diagram for 1 loop fermions phi}
\end{equation}
In the low energy EFT, this generates  $\mathcal{L}_\text{dCS}=\alpha\,\mpl\,\phi\,^*RR/4$ with coefficient (see Refs.~\cite{Toms:2018wpy,Alexander:2022cow,Alviani:2025xvf} and Appendix \ref{app: UV Completion of dCS and EFT via Heat Kernel}) given by
\begin{equation}
\alpha\,\mpl=\sum_n{g_n\over 12(4\pi)^2 m_n}.
\end{equation}
For simplicity, let us consider the case in which all $N$ fermions have the same mass $m_\psi$ and coupling $g$. 
In fact the above UV completion provided by $\Phi$ demands that $g_n/m_n=1/f$ is fixed for all $n$.
Then the coefficient is
\begin{equation}
\alpha\,\mpl={g\,N\over 12(4\pi)^2 m_\psi}.
\label{alphaUV}\end{equation}
This suggests that we can make the dCS coefficient arbitrarily large at fixed fermion mass $m_\psi$ by simply making $g$ or $N$ large. As this UV completion appears entirely causal, this would seem to contradict our above causality bound in Eq.~(\ref{c03}) which says we can bound $\alpha$ by $\alpha\sim 1/\Lamc^2<1/m_\psi^2$, and cannot be arbitrarily large.
However, we cannot in fact make $\alpha$ arbitrarily large in this construction, as we now explain. 


\subsection{Coupling and Species Bounds}

Firstly, consider a process in which we scatter a pair of (relativistic) fermions off one another via $\phi$ exchange. At tree-level the scattering amplitude is schematically
\begin{equation}
\mathcal{A}_\text{tree}=\SChannelAxion[0.5]\sim\,g^2.
\end{equation}
By allowing the exchanged scalar to have a 1-loop fermion insertion, we have (up to log corrections)
\begin{equation}
    \mathcal{A}_\text{1-loop}
    =\SChannelAxionLoop[0.5]
    \sim\,{N\,g^2\over(4\pi)^2},
\end{equation}
and we can go on in a similar manner to higher loops. By demanding that the higher order terms in the loop expansion  are not larger than the lower order terms, so that the theory remains perturbative, we have the inequality
\begin{equation}
    \lambda \equiv  {N\,g^2\over(4\pi)^2}<1. \label{lambdavalue}
\end{equation}
So although we can imagine the number of fermions $N$ being large, we have to self-consistently make the coupling $g$ small to remain perturbative  
(this is a weakly coupled  regime of a kind of 't Hooft coupling \cite{tHooft:1973alw}).

Secondly, consider a process in which we scatter any pair of particles (such as Standard Model particles) off one another via graviton exchange. At tree-level the scattering amplitude is (not including the $t$- and $u$-channel diagrams)
\begin{equation}
    \mathcal{A}_\text{tree} = \SChannelGraviton[0.5]\sim{E^2\over\mpl^2}.
   \end{equation}
At next order we can insert the above $N$ fermions in a loop from the exchanged graviton. The amplitude is
\begin{equation}
\mathcal{A}_\text{1-loop} = \SChannelGravitonLoop[0.5]  \sim{N\,E^4\over(4\pi)^2\mpl^4}.
    \end{equation}      
So even though we may be studying the scattering of Standard Model particles, these decoupled fermions can  alter the scattering amplitude since gravity is universal.

In order for the higher loop corrections to remain lower than the leading order terms, we obtain a type of ``species bound'' 
\begin{equation}
    \Lams=b{4\pi\,\mpl\over\sqrt{N}}.
\label{LN}\end{equation}
Here we have included an $\mathcal{O}(1)$ factor $b$.
We note that while the above perturbative argument suggests $b\sim 1$, there are other arguments in the literature based on black holes that suggest a slightly lower bound with $b\sim1/(4\pi)$ \cite{Dvali:2007hz}.
We note that due to this scale $\Lams$, it is thought that this so-called ``UV completion'' is only raising the energy scale of validity of the effective theory, but it still remains only an effective theory.

Now we use Eq.~(\ref{LN}) to eliminate $N$ in favor of $\Lams$, we then use Eq.~(\ref{lambdavalue}) to eliminate $g$ in favor of $\lambda$, and we insert both into the expression for the dCS coupling in Eq.~(\ref{alphaUV}) to give
\begin{equation}
    |\alpha|={b\sqrt{\lambda}\over12\, m_\psi\,\Lams}.
\label{alphaSpecies}\end{equation}

By then re-writing the dCS coupling $\alpha$ in terms of the scale being probed $L$ (\ref{alphadL}), and again expressing the result in terms of $\delta$, we obtain
\begin{equation}
\delta={b^2\over144}\,(\lambda)\,
\left(1\over L\,m_\psi\right)^2
\left(1\over L\,\Lams\right)^2.
\label{Lm}\end{equation}
The species scale $\Lams$ is plausibly a scale in which radical new physics enters (perhaps extra dimensions or strings or a breakdown of QFT, etc). It is sometimes thought that $\Lams^{-1}$ should in fact be smaller than microscopic scales probed at colliders $\sim 10^{-19}$\,m, or at least the scales probed in tabletop searches for deviations from Newtonian gravity  $\sim 10^{-4}$\,m.
So it is normally thought that we need $\Lams^{-1}$ to be very tiny compared to the usual macroscopic scales in astrophysical processes, i.e.,
\begin{equation}
    L\gg \Lams^{-1}.
\end{equation}
Then, the last term in brackets in (\ref{Lm}) should be very, very small.
Also, we have already mentioned above that $L>m_\psi^{-1}$, so the third term in brackets must be small. We also have the rescaled coupling $\lambda<1$, as discussed above. Finally, the factor of $b^2/144$ is $\sim 10^{-2}$ for $b\sim 1$ or $10^{-4}$ for $b\sim 1/(4\pi)$, and is also small. 

Altogether, the product on the right hand side of Eq.~(\ref{Lm}) implies that $\delta$ must be extremely tiny. This suggests that the detectability of the dCS operator on macroscopic dynamics is very limited. A useful representation of this result is given ahead in Eq.~(\ref{deltaEst}), which demonstrates this point more clearly.

\subsection{Other Operators}


There are other effects that arise from the above UV completion in Eq.~(\ref{eq: uv completion action of dCS}). 
For example, we can consider the following diagram involving 2 external $\phi$ particles
\begin{equation}
 \BoxAAtoHHLoop[0.5]\label{eq: feynman diagram for 1 loop fermions}
\end{equation}
As discussed in Appendix \ref{app: UV Completion of dCS and EFT via Heat Kernel}, this generates higher order operators that also contribute to the effective action, including
\begin{equation}
    \Delta\mathcal{L}= \sum_n{d_1\over(4\pi)^2}\frac{g_n^2}{m_n^2}\,\phi^2\,R_{abcd}R^{abcd}\label{eq: non vanishing EFT term propto fermion mass}.
\end{equation}
Here, $d_1$ is an $\mathcal{O}(1)$ coefficient, whose precise value is scheme dependent. 
For all masses and couplings equal, let us estimate the size of this as
\begin{equation}
    \Delta\mathcal{L}\sim {N\,g^2\over (4\pi)^2\,m_\psi^2}{\mpl^2\over L^4}\,\delta
\end{equation}
where we estimated $R\sim 1/L^2$, and used Eqs.~(\ref{phiEst}) and (\ref{alphadL}) to estimate $\phi\sim\sqrt{\delta}\,\mpl$.
By comparison, the dCS term can be estimated as
\begin{equation}
    \mathcal{L}_\text{dCS}={\alpha\,\mpl\over4}\phi\,^*R\,R\sim
    {N\,g\over (4\pi)^2\,m_\psi}{\mpl\over L^4}\,\sqrt{\delta}.
\end{equation}
The ratio of these expressions is
\begin{equation}
    {\Delta\mathcal{L}\over\mathcal{L}_\text{dCS}}
    \sim{g\,\mpl\over m_\psi}\sqrt{\delta}
    \sim \sqrt{\lambda}\,{\Lams\over m_\psi}\sqrt{\delta}.
\end{equation}
Then using the expression for $\delta$ in (\ref{Lm})  we obtain
\begin{eqnarray}
        {\Delta\mathcal{L}\over\mathcal{L}_\text{dCS}}
        \sim {\lambda\over(L\,m_\psi)^2}.\label{eq: ratio of eft correction to dCS}
\end{eqnarray}
Since the dCS theory does not include these higher order terms $\Delta\mathcal{L}$, we therefore need this to be small. So another bound for the validity of the dCS effective theory is
\begin{equation}
    {\lambda\over(L\,m_\psi)^2}\ll1.
\end{equation}
This is not too surprising as we already explained that $\lambda<1$ and $L\,m_\psi>1$. But this emphasizes that this ratio should be quite small; we cannot merely saturate these bounds and ignore higher order operators.
This implies that the product of the second and third factors in (\ref{Lm}) should be quite small. 

The above term breaks the shift symmetry in $\phi$ and is therefore in fact absent if we use the UV completion provided in Eq.~(\ref{eq: Phi uv completion action of dCS}). 
Nevertheless there are other shift symmetry obeying operators that are generated. Let us return to the processes in (\ref{eq: feynman diagram for 1 loop fermions phi}) describing  $\phi\to h h$. In addition to the dCS term, one also create a tower of other terms suppressed by derivatives, including
\begin{equation}
\Delta\mathcal{L}'= \sum_n{e_1\over(4\pi)^2}{g_n\over m_n^3}\nabla\nabla\phi\,^*R\,R
\end{equation}
where $e_1$ is an $\mathcal{O}(1)$ arising from a form factor.
The ratio of this to the dCS term is
\begin{equation}
    {\Delta\mathcal{L}'\over\mathcal{L}_\text{dCS}}\sim{1\over(m_\psi\,L)^2}
\end{equation}
This shows that we need $L\,m_\psi\gg 1$. So not merely saturating the bound.

All this stresses that the entire expression for $\delta$ needs to be very small, and once again that dCS dynamics can be very difficult to observe on macroscopic scales.

\section{Conclusions and Discussion}\label{sec: conclusions and discussion}

In this paper, motivated by causality bounds on extensions of GR, we derived a dispersion relation for dCS gravity on a background where $\overline{R}_{ab} = 0$, and then specifically considered a gravitational wave background. Along this background, we computed the Shapiro time delay to both first and second order in the GR term and first order in the dCS time delay term. We found that the first order GR term was highly suppressed in the wide Gaussian limit, requiring us to work to second order in the metric $\bar{h}_{ab}$. 
By combining all of these pieces, and demanding the time advance is no larger than the resolution of observation, we found  a causality bound on $\alpha$. This became more and more restrictive in the limit in which the frequency of the wave packet of the perturbation is taken larger and larger. It is often thought that this limit is required to examine causality, which would drive $\alpha\to0$. By instead only allowing for finite frequencies, since the EFT is expected to breakdown in the UV, we find a result that is $\Lamc\sim 0.5/\sqrt{|\alpha|}$; moderately sharper than the existing literature, but comparable.

We then examined this in the context of a known UV completion from integrating out a set of massive chiral fermions. Perturbative unitarity and a gravitational species bound imply a potentially much more extreme bound on the dCS coupling, depending on the assumption of the species cutoff $\Lams$. 

By expressing the dCS coupling $\alpha$ in terms of $\delta$, an estimate for the fractional correction to the dynamics, and $L$, a length scale of interest, we obtained in Eqs.~(\ref{alphaSpecies},\,\ref{Lm}) the result
\begin{equation}
    \delta\equiv {\alpha^2\over L^4}
    \approx 6\times 10^{-20}\left(b\over 1\right)^2\left(\lambda\over 1\right)\left(1\over L\,m_\psi\right)^2\left(10\,\text{km}\over L\right)^2\left(\Lams^{-1}\over 30\,\mu\text{m}\right)^2.
\label{deltaEst}\end{equation}
Here we have scaled our quantities by some suggestive representative values: For astrophysical black holes of the sort LIGO/Virgo can analyze, the characteristic horizon scale is order 10's of km, so we normalized $L$ to 10\,km as a useful reference value. Also, given that we have not seen any new behavior of gravity down to $\sim30\,\mu$m from tabletop tests of gravity, we normalized the species bound to this value. Recall that we have $b\lesssim1$ and $\lambda<1$ and we need $L\,m_\psi\gg1$, so all these factors are smaller than 1. This leads to an upper bound on $\delta$ on the order $10^{-20}$. Furthermore, 
one might impose that the species bound $\Lams^{-1}$ is in fact smaller than the scales probed at colliders $\sim 10^{-19}\,m$, which would reduce this upper bound significantly further to $\sim 10^{-57}$. In short, for any microscopic choice for the species bound (at which radical new physics is thought to enter)  $\delta$ and $\alpha$ are extremely small. Let us also note that if one adopts a mass value of known fermions as a guide, say from neutrinos of $m_\psi\sim 0.05$\,eV, then this suppresses this even further.

Quite a few future directions are possible. First would be to perform a more general analysis of gravitational wave backgrounds to check on a slightly sharper causality bound. This would also be interesting in its own right to compute time delay effects to second order not computed prior. 
The second direction would be to re-create the second order correction here to the (Shapiro) time delay in the amplitudes language that was seen in \cite{Alexander:2025gdn,Serra:2022pzl}. Specifically, one should identify what Feynman diagrams, or specific summation of them, gives the correct higher order corrections to the Shapiro time delay. 
Third, it would be interesting to determine what other operators could generate parity violations in the gravity sector other than dCS.
We suspect there could be a stronger statement made about parity violation if a more systematic approach was taken via either time delay considerations or via analyticity of scattering amplitudes. 

Also, one should check for configurations in which the higher order operators are surprisingly large, compared to the simple power counting estimates above where we wrote $|\alpha|=\sqrt{\delta}\,L^2$ and indicated $\delta$ is the fractional change on a length scale $L$ (for a parity violating configuration, such as a Kerr black hole). An example in which enhancement can occur is given in Ref.~\cite{Horowitz:2024dch}. However, any reasonable enhancement above the simple power counting is unlikely to overcome the extremely small estimate summarized in Eq.~(\ref{deltaEst}) from the chiral fermion UV completion. Examining other possible UV completions is therefore of interest. 

Finally, adapting our analysis to the early universe and inflation in which the length scales $L$ involved are much, much smaller can potentially overcome the extreme numbers seen above. An investigation into dCS during inflation and the early universe is left for future work.



\acknowledgments

M. P. H. is supported in part by National Science Foundation grant PHY-2310572. A. P. C. would like to thank Ciaran McCulloch for discussions on implications of parity-odd operators in cosmological correlators, and Alize Sucsuzer for discussions on time delay near black holes.
We also thank Soubhik Kumar and Grant Remmen for discussion.

\appendix
\counterwithin*{equation}{section}
\renewcommand\theequation{\thesection\arabic{equation}}

\section{Perturbed Equations of Motion}\label{sec: appendix for perturbed eom and consistency}

In this appendix, we derive the perturbed equations of motion, and then show explicitly the solution for the dispersion relation is self-consistent. By self-consistent, we mean that regardless if you eliminate $\delta\phi$ or $h_{ab}$ in the equations of motion, you will find the same dispersion relation. For the construction and usage of the TT-gauge/TT-space, please see \cite{York:1974psa,Tsagas:2007yx,Maggiore:2007ulw,Figueroa:2011ye,Dai:2012bc} and references therein.

\subsection{Perturbed Equations of Motion in Detail}\label{subsec: appendix for perturbed eom}

The equations of motion with respect to the metric and the scalar field are
\begin{align}
   \mpl^2\, G^{ab} - &\frac{1}{2}\nabla^a\phi\nabla^b\phi + \frac{1}{4} g^{ab}\left[(\nabla_c\phi)(\nabla^c\phi) + 2V(\phi)\right]\nonumber\\
    &+ \alpha M_\text{Pl}\left({}^*R^{(a)cd(b)}\nabla_c\nabla_d\phi + (\nabla_c\phi)\epsilon^{cde(b}\nabla_eR^{a)}{}_d\right)= 0\label{eq: appendix metric eom}\\
    & + \nabla_a\nabla^a\phi - V'(\phi)
    +\frac{\alpha}{4}{}^*R^{cd}{}_{a}{}^bR^a{}_{bcd} = 0\label{eq: appendix scalar eom}
\end{align}
where we have included a scalar potential for generality, the symmetrized operation is defined as $A^{(ab)}\equiv\frac{1}{2}(A^{ab} + A^{ba})$\footnote{The anti-symmetric indices are defined similarly, $A^{[ab]}\equiv\frac{1}{2}(A^{ab}-A^{ba})$.} and  $\mpl^2 = (16\pi G)^{-1}$. As explained above, we perturb the equations of motion to first order as $g_{ab} = \overline{g}_{ab} + h_{ab}$ and $\phi = \overline{\phi} +\delta\phi$ for the metric and scalar respectively, where any quantity with a bar, $\overline{(.)}$ denotes the background value. For our purposes, we choose the background $\overline{R}_{ab} = 0$ to look at gravitational wave backgrounds, and we take  $\overline{\phi} = 0$ as well as the background Pontryagin density vanishes. This simplifies our equations of motion greatly and reduces to the following,
\begin{align}
    \frac{\alpha M_\text{Pl}}{2}\left({}^*\overline{R}_{bcde}\overline{R}_a{}^{cde}h^{ab} - {}^*\overline{R}_{debc}\overline{R}_a{}^{cde}h^{ab} + {}^*\overline{R}_{adbc}\overline{\nabla}^d\overline{\nabla}^ch^{ab}\right) + \overline{\nabla}_a\overline{\nabla}^a\delta\phi& =0\label{eq: appendix perturbed scalar eom}\\
    \frac{\mpl^2}{2}\overline{\nabla}^c\overline{\nabla}_ch^{ab} - \frac{\alpha M_\text{Pl}}{4}\left({}^*\overline{R}^a{}_c{}^b{}_d\overline{\nabla}^c\overline{\nabla}^d\delta\phi + 2{}^*\overline{R}^b{}_c{}^a{}_d\overline{\nabla}^c\overline{\nabla}^d\delta\phi\right)& = 0\label{eq: appendix perturbed metric eom}
\end{align}
The perturbed metric equation of motion \eqref{eq: appendix perturbed metric eom} can be simplified by recognizing the two terms multiplied by $\alpha$ are symmetric in $(ab)$, and since the metric perturbation is symmetric in $(ab)$ as well, \eqref{eq: appendix perturbed metric eom} can be rewritten as 
\begin{equation}
    \mpl^2\,\overline{\nabla}^c\overline{\nabla}_ch^{ab} = 2\alpha M_\text{Pl}{}^*\overline{R}^b{}_c{}^a{}_d\nabla^c\nabla^d\delta\phi\label{eq: appendix final metric eom}
\end{equation}
which is the quoted equation of motion stated above. The perturbed scalar equation of motion \eqref{eq: appendix perturbed scalar eom} can also be reduced in a similar manner. Notice that the second term in $\alpha$-parenthesis can be written as $\frac{1}{2}\overline{R}_a{}^{cde}{}^*\overline{R}_{debc} = -\frac{1}{2}R_a{}^{cde}{}^*R_{bcde}$. The two terms can be added, and then we are left with
\begin{equation}
    \overline{\nabla}_a\overline{\nabla}^a\delta\phi = -\alpha M_\text{Pl}\left(\overline{R}_a{}^{cde}{}^*\overline{R}_{bcde}h^{ab} + {}^*\overline{R}_{adbc}\overline\nabla^d\overline{\nabla}^ch^{ab}\right)\label{eq: appendix final scalar eom}.
\end{equation}

We now proceed to Fourier transform the equations of motion with the covariant derivatives becoming $\nabla_a\rightarrow ik_a$ while the perturbations become
\begin{equation}
    h_{ab} = \sum_s\int_k\varepsilon_{ab}^{(s)}(k)h_s(k)e^{ik\cdot x}\:\:\:\text{and}\:\:\:\delta\phi = \int_k\varphi_ke^{ik\cdot x}\label{eq: mode expansion for graviton and scalar}
\end{equation}
where $\int_k\equiv\int d^4k/(2\pi)^4$. The perturbed equations of motion \eqref{eq: appendix final metric eom} and \eqref{eq: appendix final scalar eom} are then
\begin{align}
    \sum_s\varepsilon_{(s)}^{ab}h_s(k) & = \frac{2\alpha }{M_\text{Pl} k^2}{}^*\overline{R}^b{}_c{}^a{}_dk^ck^d\varphi_k\label{eq: app fourier metric eom}\\
    k^2\varphi_k & = \alpha M_\text{Pl}\left(\overline{R}_a{}^{cde}{}^*\overline{R}_{bcde} - k^ck^d{}^*\overline{R}_{adbc}\right)\sum_s\varepsilon_{(s)}^{ab}h_s(k)\label{eq: app fourier scalar eom}.
\end{align}
If we substitute \eqref{eq: app fourier metric eom} into \eqref{eq: app fourier scalar eom} then we find
\begin{equation}
    k^4 = -2\alpha^2\left(\overline{R}_a{}^{cde}{}^*\overline{R}_{bcde}{}^*\overline{R}^b{}_e{}^a{}_f - {}^*\overline{R}_{adbc}{}^*\overline{R}^b{}_e{}^a{}_fk^dk^c\right)k^ek^f.\label{eq: app R^3 term}
\end{equation}
The first term in \eqref{eq: app R^3 term} vanishes on $\overline{R}_{ab} = 0$ backgrounds, for which we argue for now. 

In the background where $\overline{R}_{ab} = 0$, we can consider the following tensor contraction of a Weyl tensor and the corresponding dual
\begin{equation}
    W_a{}^{cde}{}^*W_{bcde} = \frac{1}{4}g_{ab}(W\cdot{}^*W).
\end{equation}
There is no other rank-2 tensor that can be built from two Weyl tensors with one Hodge dual in $D=4$ \cite{Edgar:2001vv} (this is a direct analog of the Maxwell identity $F_{ac}{}^*F_b{}^c = \frac{1}{4}g_{ab}F\cdot{}^*F$). The remaining tensors multiplying then are symmetric and tracefree in the $ab$-indices, and so the contraction vanishes \cite{Edgar:2001vv}. Thus we are left with
\begin{equation}
    k^4 = 2\alpha^2{}^*\overline{R}^b{}_{ef}{}^a{}^*\overline{R}_{dabc}k^ck^dk^ek^f\label{eq: app dispersion relation from scalar}
\end{equation}
which matches equation (11) in \cite{Garfinkle:2010zx}. However, what has not been shown is whether or not if we take \eqref{eq: appendix perturbed scalar eom} and substitute into \eqref{eq: appendix perturbed metric eom}, and then trace over with $\varepsilon_{ab}^{(s)}$ gives the same dispersion relation. We show this in the next subsection.

\subsection{Self-Consistency of Dispersion Relation in TT-Gauge}\label{subsec: self-consistency of dispersion relation}

If we substitute \eqref{eq: appendix perturbed scalar eom} into \eqref{eq: appendix perturbed metric eom}, then we find
\begin{equation}
    k^4\sum_s\varepsilon_{(s)}^{ab}h_s(k) - 2\alpha^2\left(\overline{R}_f{}^{hij}{}^*\overline{R}_{ghij} - {}^*\overline{R}_{figh}k^hk^i\right){}^*\overline{R}^b{}_c{}^a{}_dk^ck^d\left(\sum_s\varepsilon_{(s)}^{fg}h_s(k)\right) = 0.\label{eq: app scalar into metric start}
\end{equation}
We can clean up \eqref{eq: app scalar into metric start} by first defining $\xi\equiv 2\alpha^2$, and define the following tensor
\begin{equation}
    T^{ab}{}_{fg}(k,\overline{R})\equiv\left(\overline{R}_f{}^{hij}{}^*\overline{R}_{ghij} - {}^*\overline{R}_{figh}k^hk^i\right){}^*\overline{R}^b{}_c{}^a{}_dk^ck^d.
\end{equation}
Notice that we can rewrite the polarization tensor in the first term as $\varepsilon_{(s)}^{ab} = \delta^{ab}{}_{fg}\varepsilon_{(s)}^{fg}$ where $\delta^{ab}{}_{fg}\equiv \frac{1}{2}(\delta^a_f\delta^b_g + \delta^a_g\delta^b_f)$. Then, \eqref{eq: app scalar into metric start} can be rewritten as
\begin{equation}
    \left(k^4\delta^{ab}{}_{fg} - \xi T^{ab}{}_{fg}\right)\sum_s\varepsilon_{(s)}^{fg}h_s(k) = 0.\label{eq: app second step in disp relation}
\end{equation}
Now, \eqref{eq: app second step in disp relation} is an operator-like equation acting on the polarization tensor. Recall that we are working in a specific setup, namely that $\overline{R}_{ab} = 0$, and on a gravitational wave background where solutions obey the transverse-traceless conditions (TT), which is the space of solutions we wish to probe. The polarization tensor obeys $k^a\varepsilon_{ab}^{(s)} = \overline{g}^{ab}\varepsilon_{ab}^{(s)} = 0$, and $\varepsilon_{ab}^{(s)}\varepsilon_{(s')}^{ab}{}^* = \delta_{s,s'}$. For massless spin-2 particles, the helicity projector under TT gauge can be written as
\begin{align}
    P_{ab,cd}(k) & = \sum_s\varepsilon_{ab}^{(s)}\varepsilon_{cd}^{(s)}{}^*\nonumber\\
    & = \frac{1}{2}(\theta_{ac}\theta_{bd} + \theta_{ad}\theta_{bc}) - \frac{1}{2}\theta_{ab}\theta_{cd}\label{eq: app spin 2 spin sum}
\end{align}
where
\begin{equation}
    \theta_{ab} = g_{ab} - \frac{1}{k\cdot r}(k_ar_b + k_br_a)\label{eq: app part of spin sum}
\end{equation}
where $r^2 = 0$ is a null vector and $k\cdot r\neq 0$. Here, the TT-projector is built from the leading-order geometric optics wavevector $k_a$ which in pure GR is $k^2 = 0$ while the dCS correction shifts it to $k^2 \neq 0$ only at higher order, after projection onto this zeroth-order TT-subspace \cite{Shore:2002gn,Shore:2007um}. Notice that $P^{ab}{}_{,ab}  =2$, which simply counts the number of physical modes. Now, given \eqref{eq: app second step in disp relation}, let us project this equation on the $TT$-basis we wish to probe by namely
\begin{align}
    P_{cd,ab}\left(k^4\delta^{ab}{}_{fg} - \xi T^{ab}{}_{fg}\right)P^{fg}{}_{,ij}h^{ij} & = 0\Leftrightarrow\nonumber\\
    \left(k^4 P_{cd,ij} - \xi(P\cdot T\cdot P)_{cd,ij}\right)h^{ij} & = 0\Leftrightarrow\nonumber\\
    \left(k^4 I_{TT} - \xi T_{TT}\right)h_{TT} & = 0
\end{align}
where we have implemented matrix notation in the last step for simplicity, $Ph = h$, $P^2 = P$, $P\varepsilon^{(s)} = \varepsilon^{(s)}$, and defined $T_{TT} = P\cdot T\cdot P$, which is a matrix that is projected onto the states that obey the $TT$-gauge (hence, $h_{TT}$ even has it since we have already imposed the TT-gauge onto it). 

Let us now go back to \eqref{eq: app second step in disp relation} and multiply from the left hand side (LHS) $\varepsilon_{ab}^{(s')*}$ which gives
\begin{equation}
    k^4\sum_s\left(\varepsilon_{ab}^{(s')*}\delta^{ab}{}_{fg}\varepsilon^{fg}_{(s)}\right)h_s(k) = \xi\sum_s\left(\varepsilon_{ab}^{(s')*}T^{ab}{}_{fg}\varepsilon^{fg}_{(s)}\right)h_s(k).
\end{equation}
The polarization tensors on the LHS can simplified to $\delta_{s',s}$ while on the RHS, we can define the matrix
\begin{equation}
    M_{s',s}(k,R)\equiv \varepsilon_{ab}^{(s')*}T^{ab}{}_{fg}\varepsilon^{fg}_{(s)}\label{eq: app matrix M}
\end{equation}
which allows us to write the dispersion relation in a compact form
\begin{equation}
    k^4h_{s'}(k) = \xi\sum_sM_{s',s}h_s(k).\label{eq: app matrix form of dispersion relation}
\end{equation}
With this form, we can immediately see that the matrix $M_{s',s}$ is $M_{s',s} = \varepsilon^{(s')*}\cdot (PTP)\cdot \varepsilon^{(s)}$, which is an eigenvalue problem. To make progress, we can use the fact that we are working with a parity-odd curvature operator and write a parity-odd operator of the 2D space of all TT-states with helicity operator $\mathcal{J}_{ab,cd}$ that acts on rank-2 tensors $X_{ab}$ as
\begin{equation}
    (\mathcal{J}\cdot X)_{ab}\equiv \frac{i}{2}\left(\epsilon^\perp{}_a{}^cX_{cb} + \epsilon^\perp{}_b{}^cX_{ca}\right),
\end{equation}
and the kernel form of the helicity operator is \cite{Dai:2012bc}
\begin{equation}
    \mathcal{J}_{ab,cd} = \frac{i}{2}\left(\epsilon^\perp_{ac}\theta_{bd} + \epsilon^\perp_{bc}\theta_{ad} + \epsilon^\perp_{ad}\theta_{bc} + \epsilon^\perp_{bd}\theta_{ac}\right)\label{eq: app helicity operator}
\end{equation}
where $\epsilon_{ab}^\perp$ is the antisymmetric 2-form which acts on the transverse 2-plane defined as
\begin{equation}
    \epsilon_{ab}^\perp\equiv \frac{\epsilon_{abcd}k^cr^d}{k\cdot r}\label{eq: antisymmetric 2-form},
\end{equation}
and the helicity operator maps $\mathcal{J}:\text{TT}\rightarrow \text{TT}$, is hermitian on the TT inner product, and $\mathcal{J}^2 = I_{TT}$, thus implying that $\mathcal{J}$ has eigenvalues $\mathcal{S}\equiv \pm1 $, namely $\mathcal{J}\varepsilon^{(s)} = \mathcal{S}\varepsilon^{(s)}$\footnote{Note that $\mathcal{S}$ denotes the sign of the helicity but $s = \pm 2$ denotes the spin. }.

Now, given that $T_{TT} = PTP$, and that
\begin{equation}
    (T_{TT}\cdot X)_{ab} = P_{ab,cd}T^{cd}{}_{ef}P^{ef}{}_{gh}X^{gh}.
\end{equation}
For the projected operator that is relevant here, allow an operator $\mathcal{O}$ in our restricted TT-space to be written as (with the help of \eqref{eq: app helicity operator}) as $\mathcal{O} = aI_{TT} + b\mathcal{J}$,
where $I_{TT}$ is the identity on $TT$-space, and the coefficients $a$ and $b$ can be computed via
\begin{equation}
    a = \frac{1}{2}\Tr_{TT}\mathcal{O}\:\:\:\text{and}\:\:\: b = \frac{1}{2}\Tr_{TT}(\mathcal{J}\mathcal{O}).
\end{equation}
For our particular gravitational wave setup, since the background is a vacuum implying $\overline{R}_{ab} = 0$, and if we restrict further to ${}^*\overline{R}\overline{R} = 0$, then there is no way to construct a pseudoscalar to multiply the parity-odd operator $\mathcal{J}$ in the above decomposition. Therefore, $b = 0$ and we find that
\begin{equation}
    T_{TT} = \Lambda(k,R)I_{TT}
\end{equation}
where $\Lambda(k,R)$ is exactly the coefficient $a$ computed as
\begin{equation}
    \Lambda(k,R) = \frac{1}{2}\Tr_{TT}(PTP),
\end{equation}
with the following result
\begin{equation}
    \Lambda(k,R) = {}^*\overline{R}^c{}_{ab}{}^d{}^*\overline{R}_{edcf}k^ak^bk^ek^f.
\end{equation}
Now returning to $M_{s',s}$, the matrix is simplified to
\begin{equation}
    M_{s',s} = \varepsilon^{(s')*}T_{TT}\varepsilon^{(s)} = \Lambda\varepsilon^{(s')*}I_{TT}\varepsilon^{(s)} = \Lambda \delta_{s,s'}.
\end{equation}
The dispersion relation then finally becomes
\begin{equation}
    k^4 = 2\alpha^2\,{}^*\overline{R}^b{}_{ef}{}^a{}^*\overline{R}_{dabc}k^ck^dk^ek^f.
\end{equation}
We therefore have complete self-consistency when deriving the dispersion relation via either path using the equations of motion. 

\section{Expressions for Shapiro at Second Order}\label{app: shapiro expressions appendix}

The setup for computing $\omega(t)$, deriving $v(t)$, and integrating to find $\Delta T_\pm$ are completely the same, the only difference is the length of the expressions.
The angular frequency is given as
\begin{align}
    \omega_\pm(t) & = \left(8e^{\frac{(t-z)^2}{\sigma^2}}+H^2(3+t(3t^2+2tz+z^2)\overline{\omega}^2-3\cos(2(t-z)\overline{\omega}) + 6t\overline{\omega}\sin(2(t-z)\overline{\omega})\right)^{-1}\times\nonumber\\
    &\times\biggm(H^2k_z\left(3+4t(t-2z)\overline{\omega}^2-3\cos(2(t-z)\overline{\omega})+6t\overline{\omega}\sin(2(t-z)\overline{\omega})\right)\nonumber\\
    & \pm e^{\frac{(t-z)^2}{\sigma^2}}\biggm(e^{-\frac{2(t-z)^2}{\sigma^2}}\biggm(
    H^4k_z^2(3+4t(t-2z)\overline{\omega}^2 - 3\cos(2(t-z)\overline{\omega})+6t\overline{\omega}\sin(2(t-z)\overline{\omega}))^2\nonumber\\
    & + \left(8e^{\frac{(t-z)^2}{\sigma^2}} + H^2(3+2(3t^2+2tz+z^2)\overline{\omega}^2-3\cos(2(t-z)\overline{\omega}) + 6t\overline{\omega}\sin(2(t-z)\overline{\omega})\right)\times\nonumber\\
    &\times\biggm(8e^{\frac{(t-z)^2}{\sigma^2}}|k|^2-8e^{\frac{(t-z)^2}{2\sigma^2}}H\biggm((k_z^2+2k_xk_y - k_y^2)\cos((t-z)\overline{\omega}) + H^2(k_x^2(11+2(t-z)^2\overline{\omega}^2)\nonumber\\
    & + k_y^2(11+2(T-z)^2\overline{\omega}^2) + k_z^2(-3-2(3t^2-2tz+z^2)\overline{\omega}^2)\nonumber\\
    &+ (k_x^2-8k_xk_y+9k_y^2 + 3k_z^2)\cos(2(t-z)\overline{\omega})\nonumber\\
    & - 6k_z^2t\overline{\omega}\sin(2(t-z)\overline{\omega})\biggm)\biggm)\biggm)\biggm)^{\frac{1}{2}}\biggm)
\end{align}
The total velocity is
\begin{align}
    v_\pm & = 1-\frac{H}{2|k|^2}e^{-\frac{(t-z)^2}{2\sigma^2}}\left(kx^2 - ky^2 + 2k_xk_y\right)\nonumber\\
    & +\frac{H^2e^{-\frac{(t-z)^2}{\sigma^2}}}{16|k|^4}\biggm(-4(5+7\cos(2(t-z)\overline{\omega})k_z^3k_y + 4k_xk_y\left(5k_y^2+\cos(2(t-z)\overline{\omega})\left[3k_y^2-2k_z^2\right]\right)\nonumber\\
    & + k_x^4\left[11-4t(t+2z)\overline{\omega}^2 + 7\cos(2(t-z)\overline{\omega})  -6t\overline{\omega}\sin(2(t-z)\overline{\omega})\right]\nonumber\\
    & + k_y^4\left[11-4t(t+2z)\overline{\omega}^2+15\cos(2(t-z)\overline{\omega}) - 6t\overline{\omega}\sin(2(t-z)\overline{\omega})\right]\nonumber\\
    &+2k_z^3\biggm(|k|(3+4t(t-2z)\overline{\omega}^2 - 3\cos(2(t-z)\overline{\omega}) + 6t\overline{\omega}\sin(2(t-z)\overline{\omega})\nonumber\\
    & - k_z\left(3+2(2t^2+z^2)\overline{\omega}^2 - 3\cos(2(t-z)\overline{\omega}) + 6t\overline{\omega}\sin(2(t-z)\overline{\omega})\right)\biggm)\nonumber\\
    & - 2k_y^2k_z\biggm(|k|\left(-3-4t(t-2z)\overline{\omega}^2 + 3\cos(2(t-z)\overline{\omega})-6t\overline{\omega}\sin(2(t-z)\overline{\omega})\right)\nonumber\\
    & + k_z\left(-5+2(4t^2+2tz+z^2)\overline{\omega}^2-13\cos(2(t-z)\overline{\omega}) + 9t\overline{\omega}\sin(2(t-z)\overline{\omega})\right)\biggm)\nonumber\\
    & -2k_x^2\biggm(k_y^2\left(-11+4t(t+2z)\overline{\omega}^2-11\cos(2(t-z)\overline{\omega})+6t\overline{\omega}\sin(2(t-z)\overline{\omega})\right)\nonumber\\
    & + k_z\biggm(|k|(-3-4t(t-2z)\overline{\omega}^2 + 3\cos(2(t-z)\overline{\omega}) - 6t\overline{\omega}\sin(2(t-z)\overline{\omega})\nonumber\\
    &+k_z\biggm(|k|\left(-3-4t(t-2z)\overline{\omega}^2+3\cos(2(t-z)\overline{\omega})-6t\overline{\omega}\sin(2(t-z)\overline{\omega})\right)\nonumber\\
    & + k_z\left(-5+2(4t^2+2tz+z^2)\overline{\omega}^2-9\cos(2(t-z)\overline{\omega}) + 9t\overline{\omega}\sin(2(t-z)\overline{\omega})\right)\biggm)\biggm)\biggm)\biggm).
\end{align}
Only computing the Shapiro time delay for the second order term, we find
\begin{align}
    \Delta T_{\pm}^{(2)} & = \frac{e^{-(\sigma\overline{\omega})^2}H^2\sqrt{\pi}\sigma\csc(\theta/2)^2}{128|k|^4}\biggm(
    -4\biggm((7+11e^{(\sigma\overline{\omega}))^2})k_x^4-4(t+5e^{(\sigma\overline{\omega}))^2})k_x^3k_y+(15_+11e^{(\sigma\overline{\omega}))^2})k_y^4\nonumber\\
    & + 4k_xk_y\left((3+5e^{(\sigma\overline{\omega}))^2})k_y^2-2k_z^2\right) + 2k_y^2k_z\left((13+5e^{(\sigma\overline{\omega}))^2})k_z + 3e^{(\sigma\overline{\omega}))^2}|k|\right) \nonumber\\
    &- 6k_z(k_z-|k|)\left(\left(e^{(\sigma\overline{\omega}))^2}-1\right)k_z^2 - k_z|k| - |k|^2\right)\nonumber\\
    & + 2k_x^2\left(11(1+e^{(\sigma\overline{\omega}))^2})k_y^2 + k_x(9k_z+5e^{(\sigma\overline{\omega}))^2}k_z + 3e^{(\sigma\overline{\omega}))^2}|k|)\right)\biggm)\nonumber\\
    &+ 12(k_x^2+k_y^2+2k_z(k_z-|k|))|k|^2\sigma^2\overline{\omega}^2\csc(\theta/2)^2\nonumber\\
    & +e^{(\sigma\overline{\omega}))^2}|k|^2\sigma^2\overline{\omega}^2\left(2k_x^2+2k_y^2+7k_z^2-4k_z|k|+4(k_x^2+k_y^2+2k_z|k|)\cos(\theta) + k_z^2\cos(2\theta)\right)\csc(\theta/2)^4\biggm).
\end{align}

We have reported on the case of the group speed here. But the case of the phase speed can be readily obtained from this by using the connection provided in Eq.~(\ref{groupphase}).
As described earlier, they match at the order we are working in the main text.

\section{Shapiro Time Delay Near a Black Hole}\label{app: shapiro time delay near a black hole appendix}

referring back at figure \eqref{fig:time delay bh image}, let us consider the case where the pulse that travels along $\Delta T$ is close to the black hole system. We would  expect for there to be a redshift correction that would modify \eqref{eq: total time delay computed in wide gaussian limit}. Modeling the binary black hole system as an effective Schwarzschild geometry far enough away, there is a corresponding timelike Killing vector with conserved energy $\omega_\infty$ that is related to a local observer's energy as \cite{Misner:1973prb} $\omega_\text{loc}(r) = \omega_\infty/N(r)$ with the lapse function being $N(r) = \sqrt{-g_{tt}} = (1-2M/r)^{1/2}$. The $\overline{\omega}^2$-factor then in the dCS contribution in \eqref{eq: total time delay computed in wide gaussian limit} is replaced by a Gaussian average along the ray
\begin{equation}
    \mathcal{R}_z\equiv \frac{\int_{-\infty}^\infty du\:N^{-2}(r(u))e^{-u^2/2\sigma_t^2}}{\int_{-\infty}^\infty du\:e^{-u^2/2\sigma_t^2}}
\end{equation}
where $u$ parameterizes the unperturbed trajectory of the ray while $\sigma_t$ is the temporal-width of the pulse along the ray itself.
We can consider a model of closest approach where $r(u) = \sqrt{b^2+u^2}$ with impact parameter $b$. If $b\gg 2M$ then $N^{-2}(r) = 1 + 2M/r + \mathcal{O}(M^2/r^2)$ and for a narrow pulse with $\sigma_t\ll b$ 
\begin{equation}
    \frac{1}{r(u)} = \frac{1}{b}\left(1 - \frac{u^2}{2b^2} + \mathcal{O}(u^4/b^4)\right).
\end{equation}
The Gaussian average then gives $\langle u^2\rangle = \sigma_t^2$ implying a Gaussian average contribution of 
\begin{equation}
    \mathcal{R}_z = 1 + \frac{2M}{b}\left(1 - \frac{\sigma_t^2}{2b^2} + \mathcal{O}(\sigma_t^4/b^4)\right) + \mathcal{O}(M^2/b^2). 
\end{equation}

\section{dCS  EFT from a UV Completion}\label{app: UV Completion of dCS and EFT via Heat Kernel}

In this appendix, we give a brief overview of the derivation for finding the chiral-rotation of the gravitational anomaly, and then outline how the other EFT operators that arise from integrating out the chiral fermion can be computed and their generic form. The approach we comment on uses the heat kernel method \cite{Vassilevich:2003xt,Toms:2018wpy}. 

Our starting point is the following action
\begin{equation}
    S = \int d^4x\sqrt{-g}\:\overline{\psi}\left(i\gamma^a \nabla_a - m_\psi + i\,g\,\phi\,\gamma_5\right)\psi\label{eq: uv completed dcs chiral fermion action}
\end{equation}
where the $\gamma$-matrices obey the Clifford commutation relation $\gamma_a\gamma_b + \gamma_b\gamma_a = 2g_{ab}$ which in this way obey $\gamma_a^\dagger = \gamma_a$, the $\phi$ is a scalar field. The chirality matrix is denoted as $\gamma_5$ which is hermitian, and anticommuites with the $\gamma$-matrices. The covariant derivative here, $\nabla_a$ is written as 
\begin{align}
    \nabla_a = \partial_a + \frac{1}{8}[\gamma_b,\gamma_c]\sigma_a^{bc}\label{eq: spinor covariant derivative}
\end{align}
where $\sigma_a^{bc}$ is the spin-connection \cite{Alexander:2022cow}. The total Dirac operator can then be written as
\begin{equation}
    \slashed{D} \equiv i\slashed{\nabla} - m_\psi + i\,g\,\phi\,\gamma^5\label{eq: UV dirac operator}.
\end{equation}
When we integrate out the fermions, we will get a contribution to the effective action of the form
\begin{equation}
    S_\text{eff}\supset -i\ln\det(i\slashed{\nabla} - m_\psi +i\,g\,\phi\,\gamma_5)
\end{equation}
where the effective action also contains the kinetic action of the scalar field and the Einstein-Hilbert action. The generation of the dCS operator is performed in \cite{Toms:2018wpy}, but for the case of a constant global gauge. Here we consider a \textit{local} chiral rotation which obtains a similar result, but keeps the dynamics of the scalar field evident. Consider a chiral rotation of the form $\psi = e^{-i\theta(x)\gamma_5}\psi$ and similar for the conjugate. If we require the mass terms of the UV theory to be invariant under this chiral rotation, then we must require the following to be true
\begin{equation}
    m_\psi\sin(2\theta) - g\phi\cos(2\theta) = 0
\end{equation}
which when we solve for the local rotation becomes
\begin{equation}
    \theta(x) = \frac{1}{2}\arctan(\frac{g\phi(x)}{m_\psi})\label{eq: chiral rotation angle}.
\end{equation}
Under this rotation, the mass term becomes a scalar term where we define a new mass $M(x) = \sqrt{m_\psi^2 + g^2\phi^2}$. This generates a new axial vector coupling, $V_a = \partial_a\theta$. The new Dirac operator is then 
\begin{equation}
    \slashed{D}' = i\slashed{\nabla} - M - \slashed{V}\gamma_5.
\end{equation}
Where $\theta(x)$ plays a role in that, it is this chiral rotation of the path-integral measure that generates the following deformation of the Lagrangian \cite{Toms:2018wpy,Alexander:2022cow}
\begin{equation}
    \Delta\mathcal{L} = \frac{\theta(x)}{768\pi^2}\epsilon^{de cf}R_{abde}R^{ab}{}_{cf},
\end{equation}
but expanding $\theta(x)$ as
\begin{equation}
    \theta(x) = \frac{g\phi}{2m_\psi} - \frac{g^3\phi^3}{6m_\psi^3} + \cdots
\end{equation}
gives the dCS term at first order\footnote{Note that this is just the usual result from the gravitational axial-vector current anomaly \cite{Alvarez-Gaume:1983ihn} but with a chiral rotation of the Dirac field.}
\begin{equation}
    \mathcal{L}_\text{dCS} = \frac{g\,\phi}{768m_\psi\pi^2} \,{}^*R\,R
\end{equation}
where the dual of the Riemann tensor is defined in \eqref{eq: dual riemann tensor} which matches a result in \cite{Alexander:2022cow}. This is the lowest mass-dimension operator that violates parity. 

Other operators that are parity even can be computed in the unrotated basis using the heat kernel method. We will not perform such a computation here since it requires careful renormalization to high order in the Seeley-deWitt coefficients, but we can write the operators in the EFT Lagrangian by simply counting mass-powers and vertex contributions. At next order in power counting, we can consider  $2\rightarrow 2$ scattering of scalars and gravitons, we can expect the generation of the following operators
\begin{equation}
    \mathcal{L}_\text{EFT}\supset \frac{g^2}{(4\pi)^2m_\psi^2}\left(d_1\phi^2R_{abcd}R^{abcd} + d_2\phi^2 R_{ab}R^{ab} + d_3\phi^2R^2 + d_4\phi^2\Box R + d_5 R(\nabla_a\phi)^2\right)\label{eq: scalar graviton coupling eft dim 6 pieces}
\end{equation}
The first 4 terms break the shift symmetry in $\phi$, while the final term does not; it is the latter that would survive if the theory descends from the complex $\Phi$ theory in Eq.~(\ref{eq: Phi uv completion action of dCS}).
\bibliographystyle{JHEP}
\bibliography{biblio}


\end{document}